\documentclass[prl,twocolumn,floatfix,nofootinbib,superscriptaddress]{revtex4-1}
\usepackage{graphicx}
\usepackage{subcaption}
\usepackage[fleqn]{amsmath}
\usepackage{amssymb}
\usepackage[usenames]{color}
\definecolor{goodgreen}{rgb}{0.1,0.5,0}
\definecolor{goodred}{rgb}{0.7,0,0}
\usepackage{pifont}
\usepackage{float}
\usepackage{multirow}
\usepackage{booktabs}
\usepackage[colorlinks,urlcolor=goodgreen,citecolor=blue,linkcolor=goodred]{hyperref}
\usepackage{bm}
\usepackage{mathtools}
\allowdisplaybreaks
\begin{document}
\title{Unhiding a concealed resonance by multiple Kondo transitions in a quantum dot}
\author{Aritra Lahiri}
\affiliation{Institut f\"ur Theoretische Physik, Universit\"at Regensburg, D-93040 Regensburg, Germany}
\affiliation{School of Physics and Astronomy, University of Minnesota, Minneapolis, MN 55455, USA}
\author{Tokuro Hata}
\affiliation{Department of Physics, Graduate School of Science, Osaka University, 560-0043 Osaka, Japan}
\author{Sergey Smirnov}
\affiliation{P. N. Lebedev Physical Institute of the Russian Academy of Sciences, 119991
	Moscow, Russia}
\author{\\Meydi Ferrier}
\affiliation{Department of Physics, Graduate School of Science, Osaka University, 560-0043 Osaka, Japan}
\affiliation{Laboratoire de Physique des Solides, CNRS, Univ. Paris-Sud, Universit\'e Paris Saclay, 91405 Orsay Cedex, France}
\author{Tomonori Arakawa}
\affiliation{Department of Physics, Graduate School of Science, Osaka University, 560-0043 Osaka, Japan}
\author{Michael Niklas} 
\affiliation{Institut f\"ur Theoretische Physik, Universit\"at Regensburg, D-93040 Regensburg, Germany}
\author{Magdalena Marganska} 
\affiliation{Institut f\"ur Theoretische Physik, Universit\"at Regensburg, D-93040 Regensburg, Germany}
\author{Kensuke Kobayashi} 
\email{kensuke@meso.phys.sci.osaka-u.ac.jp}
\affiliation{Department of Physics, Graduate School of Science, Osaka University, 560-0043 Osaka, Japan}
\affiliation{Center for Spintronics Research Network (CSRN), Graduate School of Engineering Science, Osaka University, Osaka 560-8531, Japan}
\affiliation{Institute for Physics of Intelligence and Department of Physics, The University of Tokyo, Tokyo 113-0033, Japan.}
	\author{Milena Grifoni} 
	\email{milena.grifoni@physik.uni-regensburg.de}
	\affiliation{Institut f\"ur Theoretische Physik, Universit\"at Regensburg, D-93040 Regensburg, Germany}
\date{\today}

\begin{abstract}
Kondo correlations are  responsible for the emergence of a zero-bias peak in the low temperature differential conductance of Coulomb blockaded quantum dots. In the presence of a global SU(2)$\otimes$SU(2) symmetry, which can be realized in carbon nanotubes, they also inhibit inelastic transitions  which preserve the Kramers pseudospins associated to the symmetry. 
We report on magnetotransport experiments on a Kondo correlated carbon nanotube  where resonant features at the bias corresponding to   the  pseudospin-{\em preserving} transitions are observed. We attribute this effect to a  simultaneous  enhancement of pseudospin-{\em non-preserving} transitions occurring at that  bias. This process is boosted by asymmetric tunneling couplings
of the two Kramers doublets to the leads and by asymmetries in the potential drops at the leads. Hence, 
the present work discloses a fundamental microscopic mechanisms ruling transport  in  Kondo systems far from equilibrium.



\end{abstract}

\maketitle

\begin{figure}[tb!]
	\begin{center}
		\includegraphics[width=\linewidth]{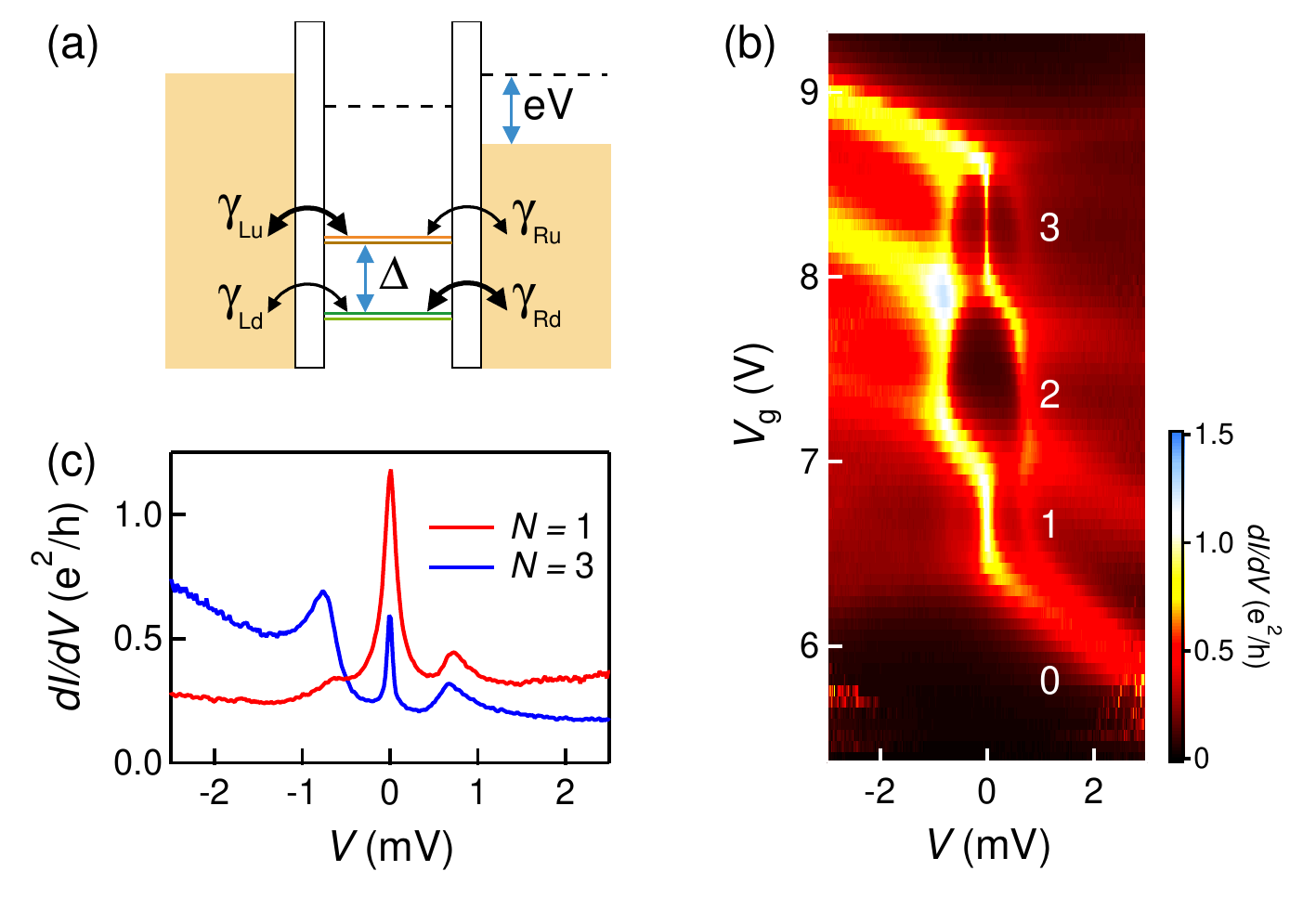}
		\caption{The SU(2)$\otimes$SU(2) Kondo effect. a) A CNT longitudinal shell  possesses two Kramers doublets separated by a splitting $\Delta$.
			The upper (u) and lower (d) doublet are coupled to left (L) and right (R) leads by tunneling couplings $\gamma_{{ L}p}$ and $\gamma_{{R}p}$, where $p={\rm u,d}$. 
			b) Experimental differential conductance as a function of gate, $V_{\rm g}$, and bias, $V_{}$, voltages for different occupation of a longitudinal shell. In the valleys with odd occupancy a zero-bias Kondo ridge is clearly seen. Two additional ridges, symmetrically located with respect to the central Kondo peak, are  observed at finite bias. The asymmetric
response implies the presence of asymmetries in the
			problem.  c) Bias traces taken at gate voltages corresponding to the center of the $N=1$ and $N=3$  valleys show  inelastic peaks with roughly the same spacing $\Delta$.  }
		\label{fig1}
	\end{center}
\end{figure}

The Kondo effect~\cite{Kondointro} is a quintessential example of strong correlations in a many-body system, stemming from the screening of a localized spin by a Fermi sea of conduction electrons. Quantum dots (QD) in the Coulomb blockade regime, which  effectively behave as a spin-1/2 system, provide a simple manifestation of SU(2) Kondo entanglement between the  dot spin  and the lead conduction electrons, leading to the formation of a many-body  spin singlet~\cite{Goldhaber1998,Cronenwett1998}. 
The Kondo effect in QDs can also have more exotic realizations, provided that the associated degrees of freedom are conserved during tunneling. A prominent example are 
carbon nanotube (CNT) QDs where 
the presence of orbital (valley) and spin degrees of freedom leads to   the  SU(4) \cite{Jarillo-Herrero2005,Choi2005,Makarovski2007PRL,Anders2008,Ferrier2016,Ferrier2017}  and the SU(2)$\otimes$SU(2) Kondo effects~\cite{Fang2008,Cleuziou2013,Schmid2015,Niklas2016}. The latter occurs when the valley and spin degeneracy of a CNT longitudinal mode is broken by spin-orbit coupling \cite{Galpin2010} or valley mixing \cite{Mantelli2016}, giving rise to two time-reversal protected Kramers doublets separated by an inter-Kramers splitting $\Delta$, as seen in Fig.~\ref{fig1}(a).   
A signature of the SU(2)$\otimes$SU(2) Kondo effect is thus the occurrence of a zero-bias anomaly accompanied by two  inelastic peaks, symmetrically located with respect to the central peak, in the differential conductance of a CNT with  single electron or single hole occupancy \cite{Schmid2015}. 
Similar features are also seen in our experiment, as shown in Figs. \ref{fig1}(b), (c).

\begin{figure}[htb!]
	\begin{center}
	\includegraphics[width=1.0\columnwidth]{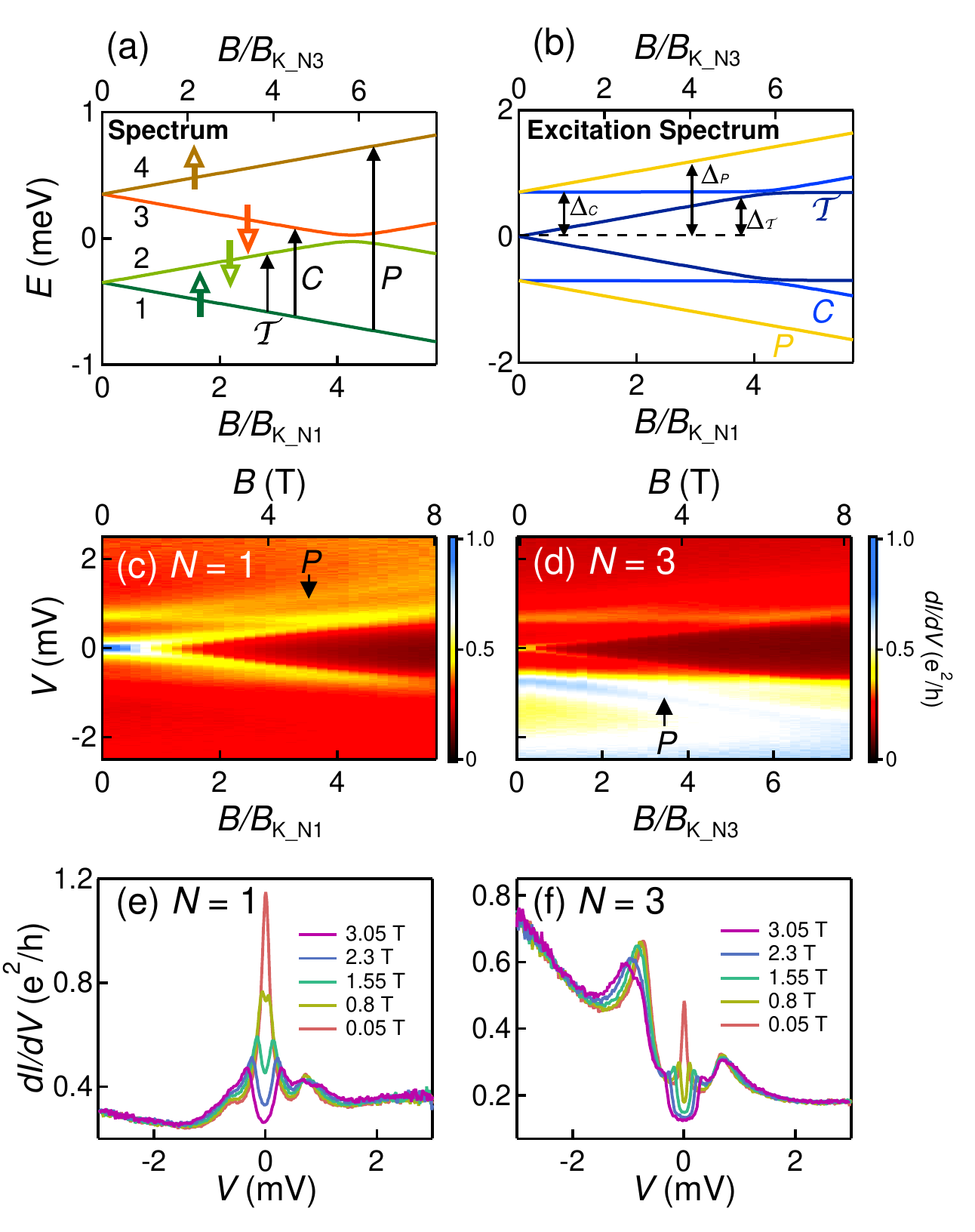}
		\caption{Behavior in perpendicular magnetic field. a) Evolution of energy levels  in magnetic field; to each level is associated a Kramers  pseudospin. b) Addition spectrum obtained from (a) with inelastic excitations $\Delta_T, \Delta_C$, and $\Delta_P$ associated to ${\cal T}$, $C$ and $P$-transitions being indicated. The magnetic field is scaled by the  characteristic   fields $B_{K_{N1}}=1.425$\,T and $B_{K_{N3}}=1.036$\,T   to emphasize universal behavior. c), d) Experimentally measured differential conductance  vs scaled  magnetic field. The $P$-like resonances are clearly visible and indicated by an arrow. (e), (f) Experimental  $dI/dV$ vs bias voltage  for various values of the magnetic field for (e) the one electron and (f) the three electron valley.  }
		\label{fig2}
	\end{center}
\end{figure}

In analogy to the more conventional SU(2) case, a pseudospin  can be associated to each Kramers doublet of the CNT and the lead electrons \cite{Mantelli2016}.
While the central peak   accounts for elastic virtual transitions which flip the pseudospin of the CNT electron within the same Kramers doublet  (${\cal T}$-transition), the inelastic peaks denote transitions involving one state in the lower and one state in the upper Kramers doublet. One distinguishes between chiral (${C}$) and particle-hole (${P}$) transitions if the two states involve the opposite or the same pseudospin, respectively (see Fig. \ref{fig2}(a)). Strikingly, inelastic transitions of the ${P}$ type are inhibited by effective anti-ferromagnetic exchange (Kondo) correlations between the pseudospins of lead and CNT electrons  \cite{Lim2006,Mantelli2016}, as confirmed by transport experiments at low magnetic fields~\cite{Jarillo-Herrero2005, Makarovski2007PRL, Cleuziou2013, Schmid2015, Niklas2016}. However, ${P}$-transitions can be observed in the weak coupling regime, where Kondo correlations do not play a role and only lowest order cotunneling processes are responsible for the inelastic peaks~\cite{Nygård2000,Jarillo-Herrero2005_PRL,Jespersen2011,Niklas2016}. 

   In this Letter, we demonstrate experimentally the puzzling emergence of a resonance at energies of the inelastic ${P}$-transition in Kondo correlated CNT QDs. Noticeably, the $P$-resonance is clearly seen only for a given bias polarity, suggesting its association with lead coupling asymmetries. We present a comprehensive theoretical analysis based on the  Keldysh effective action (KEA) theory~\cite{Schmid2015,Smirnov2013}, addressing the role of asymmetries in Kondo-correlated CNT QDs. The ${P}$-like features arise from the coherent addition of a ${C}$- and a ${\cal T}$-type transition, which occurs when the applied bias equals the energy of the inelastic ${P}$-transition, becoming relevant for different couplings of the Kramers doublets to the leads. 
%
\textit{Experimental results.-} Our device is made of a CNT grown by chemical vapour deposition and connected to Pd($6\ {\rm nm}$)/Al($70\ {\rm nm}$) leads. Fabrication details can be found in Ref. \cite{Ferrier2016}.
The differential conductance of our CNT QD is shown in Fig. \ref{fig1}(b) as a function of the applied bias $V$ and gate voltage $V_{\rm g}$. A small perpendicular magnetic field $B= 0.02$~T is applied to suppress superconductivity of Al in the leads. A  Kondo ridge, corresponding to the yellow line at zero bias, is recognized in the Coulomb valleys with occupation $N=1$ and $N=3$ of a longitudinal shell. In the $N=0, 2$  valleys, in contrast, no Kondo ridge is seen. Additional inelastic peaks, symmetrically located with respect to zero bias, are observed for the $N=1, 2, 3$ valleys. Bias traces taken at gate voltages corresponding to the center of a valley are shown in Fig. \ref{fig1}(c).  From such traces a Kramers splitting of $\Delta \simeq 0.7$~meV is estimated. Additionally, from the width of the zero-bias peaks  Kondo temperatures of $T_{K1}=1$~K and $T_{K3}=0.37$~K are extracted for valley $N=1$ and $N=3$, respectively. 
Since our experiments are taken at temperatures around $T=30$~mK, it holds  $T<T_K$. 
Furthermore, from the evolution in perpendicular magnetic field (see Eq. (\ref{deltaB}) below), we can extract a spin-orbit coupling splitting $\Delta_{\rm SO}=0.07$~meV and a larger valley mixing energy  $\Delta_{{\rm K,K'}}=0.7$~meV for both  Kondo valleys. 

A magnetic field breaks time-reversal symmetry and hence also Kramers degeneracy. The expected evolution of the single particle energy spectrum of a longitudinal shell and its associated excitation spectrum are shown in Figs. \ref{fig2}(a) and (b). The Kondo effect is however a many-body phenomenon, and differences in the excitation spectrum are expected. The experimental magnetoconductance is shown in Figs. \ref{fig2}(c) and (d); the  reference field  $B_K=eV_K/\mu_{\rm B}$ is defined through the value $V_K$ of the applied bias voltage where the low bias differential conductance  $dI/dV$ is $2/3$ of its value at  zero-bias. This ensures universal behavior of the scaled magnetoconductance \cite{Gaass2011}.
	Differential conductance traces for different  values of the magnetic field are shown in Figs. 
	\ref{fig2}(e) and (f) and characterize the behavior at low fields. We find that 
 the zero bias peak in valley $N=1$ and $N=3$ only splits above a critical field of the order of the reference field $B_K$, as expected from theoretical predictions \cite{Costi2000,Smirnov2013NJP}. Further, the inelastic peaks do not split nor move at small values of the magnetic field for valley $N=1$, suggesting a predominance of inelastic $C$-transitions \cite{Schmid2015,Niklas2016}.
Valley $N=3$ can be viewed as a shell with a single hole. Here the side peak at negative bias moves towards larger negative values of the bias voltage as the field increases, suggesting that a $P$-like transition is observed.  Hence, the behavior is strongly asymmetric in the bias voltage and the $P$-like resonance is seen only for positive (negative) voltages  for electron (hole) transport. In the following we propose a theoretical explanation for the experimental findings.

\textit{Model and KEA self-energy.-}
We consider the four-levels Anderson model to describe a longitudinal mode of a CNT quantum dot with both orbital and spin degrees of freedom. We denote by $| j\rangle$, $j=1,..,4$ the  single-particle eigenstates and associate to the lower Kramers doublet the couple (1,2), to the upper the couple (3,4), see  Fig. \ref{fig2}(a). The CNT Hamiltonian thus has the  form
\begin{equation}
\label{Hamiltonian}
\hat{H}_{\rm CNT}=\sum_{ j }\varepsilon_j (B)\hat{n}_j + \frac{U}{2}\sum_{i\neq j} \hat{n}_i \hat{n}_j +\hat H_J, 
\end{equation}
where $\hat{n}_j=\hat{d}^\dagger_j\hat{d_j}$ is the occupation operator of level $j$, and   $\varepsilon_{1,4} (B) =  \varepsilon_d \mp  \Delta(B)/2$, $ \varepsilon_{2,3} (B) =\varepsilon_{1,4} (-B)$. Here  
\begin{align}
\Delta (B)&=\sqrt{\Delta_{\mathrm{SO}}^2 + (\Delta_{\mathrm{K,K'}} + g_s\mu_B B)^2}=:\Delta_P \label{deltaB} 
\end{align} is the magnetic field dependent level splitting ($g_s\mu_B$ is the spin magnetic moment). As indicated in the last equality, such level splitting yields the addition energy for the $P$-resonance. 
Finally, the second and third terms in Eq. (\ref{Hamiltonian}) account for charging and exchange effects, respectively. We consider strong electron-electron interactions $(U=\infty)$, such that double occupancy of the impurity is excluded and exchange effects are not relevant. The evolution of the four energy levels $\varepsilon_j(B)$ in  magnetic field is shown in Fig. \ref{fig2}(a) 
 together with the possible  transitions (${\cal T}$, $C, P$) from the ground state. 
 The complete single-particle excitation spectrum is illustrated in Fig. \ref{fig2}(b).
 Notice that  $C$-excitations are independent of the magnetic field until the anticrossing of the inner levels (2,3).  Further, it holds the relation 
$\Delta_P=\Delta_T+\Delta_C$, with $\Delta_T=\varepsilon_{2}-\varepsilon_1=\varepsilon_{4}-\varepsilon_3$, and $\Delta_C=\varepsilon_3-\varepsilon_1=\varepsilon_4-\varepsilon_2$. 
 
 Kondo correlations modify the simple single-particle picture, as shown in  Figs. \ref{fig2}(c)-(f).
To account for this behavior, we have evaluated the differential conductance of a  four-levels  Anderson model with bias and tunneling asymmetries using the Keldysh effective action (KEA) method. Assuming that the Kramers degrees of freedom are conserved during tunneling \cite{Choi2005,Lim2006}, KEA yields the tunneling density of states (TDOS)
of channel $j$ \cite{Schmid2015}
\begin{equation}
\label{TDOS}
\nu_j(\varepsilon,{B})=
\frac{\Gamma/2\pi}{[\varepsilon_j({B})-\varepsilon+\Gamma_j {\rm Re}\Sigma^\text{}_j(\varepsilon,{B})]^2+[\Gamma_j {\rm{Im}}\Sigma^\text{}_j(\varepsilon,{B})]^2},
\end{equation}
in terms of the KEA self-energies $\Sigma_j={\rm{Re}}\Sigma_j+i{\rm{Im}}\Sigma_j$ being the central quantities of the theory. Here $\Gamma=\sum_{\alpha,j}\Gamma_{\alpha j}/4$ is the average coupling and $\Gamma_{\alpha j}$ are the tunneling couplings of channel $j$ at lead $\alpha=L,R$.  
The  current  follows from the Meir and Wingreen formula \cite{Meir1992}
\begin{equation}
\label{current2}
I=\frac{e}{\hbar}\sum_{j=1}^{4} \int_{-\infty}^{\infty}d\varepsilon \frac{\Gamma_{Lj}\Gamma_{Rj}}{\Gamma_{Lj}+\Gamma_{Rj}}\nu_j(\varepsilon)[f_L(\varepsilon)-f_R(\varepsilon)],
\end{equation}
where $f_\alpha=[\exp\beta(\varepsilon -\mu_\alpha)+1]^{-1}$ is the Fermi function, and $\mu_{L}=\mu_0+\eta eV_{},\mu_{R}=\mu_0-(1-\eta)(eV_{})$ with $\eta\in[0,1]$ accounting for an 
asymmetric bias drop between the left and right  leads. 
 The coupling asymmetry parameter for the lead $\alpha$ and level $j$ is given by $\gamma_{\alpha j}=\Gamma_{\alpha j}/\Gamma_j$, with $\Gamma_{j}=\sum_\alpha \Gamma_{\alpha j}$.
We keep the SU(2) symmetry within the same Kramers channel, and set
\begin{equation}
\label{symmetrySU2}
\gamma_{\alpha 1}=\gamma_{\alpha 2}:=\gamma_{\alpha \rm{d}}, \quad \gamma_{\alpha 3}=\gamma_{\alpha 4}:=\gamma_{\alpha \rm{u}},
\end{equation}
as illustrated in Fig. \ref{fig1}(a).  Such asymmetries  enter in the channel self-energies $\Sigma_j$, and hence impact the relevance of a given transition. For occupation $N=1$ we find 
\begin{equation}
\begin{split}
&\Sigma_{j }(\varepsilon,B)=\frac{1}{\pi }\sum_{  i=T_j, C_j}\frac{\Gamma_i}{\Gamma_j}\Bigg[ \ln \left( \frac{W}{2\pi k_{\rm B}T} \right) + \frac{i \pi}{2} \label{S_e2c}\\
& - \sum_{ \alpha}   \gamma_{\alpha i}\Psi\left( \frac{1}{2} +\frac{\cal E}{2\pi k_{\rm B}T} -i\frac{\mu_\alpha -\varepsilon +\Delta_{ji}}{2\pi k_{\rm B}T}   \right)\Bigg] ,
\end{split}
\end{equation}
where $W$ is a high energy cut-off,   $\Psi$ is the digamma function,  
$\Delta_{ji}=\varepsilon_j-\varepsilon_i$, and $T_j, C_j$ are the ${\cal T}$- and $C$-partners of level $j$. The case $N=3$  is obtained from Eq. (\ref{S_e2c}) upon replacement of $\Delta_{ji}\to -\Delta_{ji}$.  
Finally, the complex quantity ${\cal E}$ accounts for low energy contributions which make the self-energy finite also at zero temperature, as discussed in Sec. IV of the Supplemental Material.

\textit{Impact of asymmetries.-}
The analytic forms Eqs. (\ref{TDOS}), (\ref{current2}) allow us to analyze asymmetry effects  on the differential conductance $dI/dV_{}$. 
Before turning to highly nonequilibrium situations with  $eV\simeq\Delta_P(B)$, we focus on low energies.   An expansion of the zero temperature and zero magnetic field differential conductance $G_{\rm{diff}}$ in powers of the applied bias,
$G_{\mathrm{diff}}=G_{0} + G_{1} V_{}+ \ldots$,  yields 
\begin{figure}[htb!]
	\begin{center}
		\includegraphics[width=\linewidth]{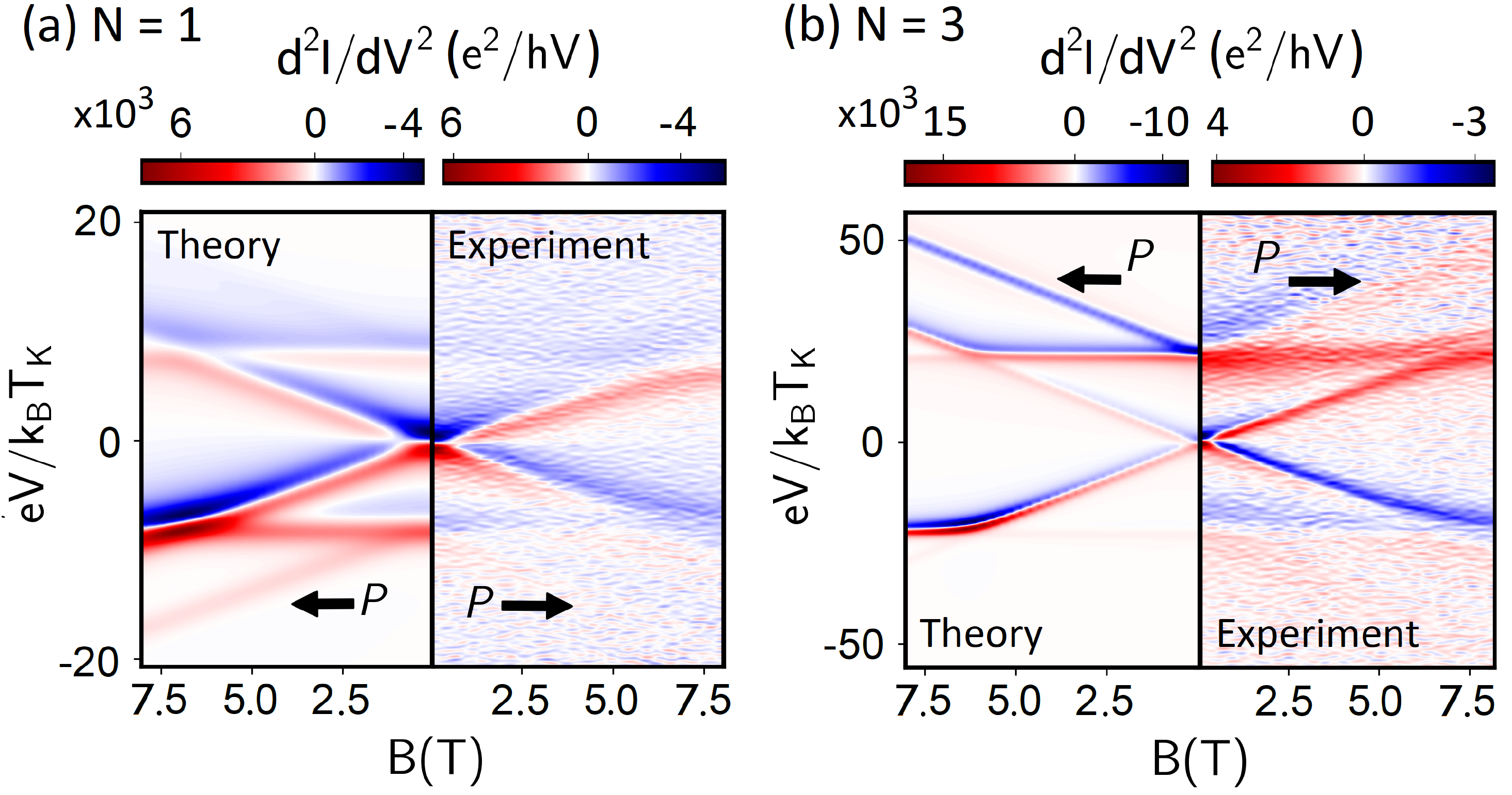}
		\caption{Second derivative of the current, $d^2I/dV_{}^2$, for the (a) $N=1$ valley and, (b) $N=3$ valley as a function of magnetic field. The left sub-panels  show analytical  predictions   for the asymmetric four-level Anderson model, the right sub-panels the experimental observations. Kondo peaks in the differential conductance are manifested as zeros of $d^2I/dV^2$ as it changes from positive (red) to negative (blue) with increasing bias. Arrows point to the $P$-resonance.  }
		\label{fig3}
	\end{center}
\end{figure}
\begin{equation}
\label{su2su2gdiff0}
G_0=\frac{2e^2}{h}\Bigg[4\gamma_{L \rm{d}}\gamma_{R \rm{d}}+
(\gamma_{L \rm{u}}\gamma_{R \rm{u}}-\gamma_{L \rm{d}}\gamma_{R \rm{d}})\frac{\pi}{2}\Gamma_{\rm u}\nu_{\rm u}(\mu_0) 
\Bigg],
\end{equation}
being  independent of  the bias asymmetry $\eta$. The second term in the bracket is proportional to the transmission of the upper Kramers doublet at the Fermi level ${\cal T}_{\rm u}(\mu_0)=\frac{\pi}{2}\Gamma_{\rm u}\nu_{\rm u}(\mu_0)$, and vanishes in the  SU(4) coupling case where $\gamma_{\alpha \rm{u}} =\gamma_{\alpha \rm{d}}=\gamma_\alpha$. Then Eq. (\ref{su2su2gdiff0}) yields the known Fermi liquid result 
$G_0=4\gamma_{L }\gamma_{R }\frac{2e^2}{h}$.
The expression for the linear term is lengthier and given in Sec. V of the Supplemental Material.
Similar to $G_0$, also $G_1$ is independent of the bias asymmetry $\eta$. Further, it is finite only 
 in the presence of lead asymmetries encapsulated in the parameter   $D_p=\gamma_{Lp}- \gamma_{Rp}$, $p={\rm u,d}$.
 For finite  $D_u=D_d$ and $\Delta=0$ we recover known results for the SU(4) case \cite{PhysRevB.80.155322}. Here the linear term $G_1$ is   non vanishing   due to a small shift of the TDOS peak from the Fermi energy, as expected from the Friedel sum rule \cite{hewson_1993}. 
These results show that asymmetries can yield {\em qualitatively} different low energy behavior of a SU(2)$\otimes$SU(2) Kondo QD with respect to the symmetric case. Further, they  suggest that the strong asymmetric behavior observed in the experimental data of Fig. \ref{fig2} requires  couplings $\gamma_{\alpha \rm{d}}\neq\gamma_{\alpha\rm{u}}$.

\textit{Resonances at finite bias.-} 
We start our analysis by showing in Fig. \ref{fig3}  KEA predictions for $d^2I/dV^2_{}$.
The parameters $\Delta_{\rm K,K'}$ and $\Delta_{\rm SO}$ are obtained from Eq. (\ref{deltaB}) by a fit of the  experimental magnetoconductance at large enough fields. The total  linewidth $\Gamma$ is extracted from a fit of the data near the charge peaks, as explained in the Supplemental Material. We fix $W/\Gamma=100$; the remaining chosen set of  free parameters  is shown in Table \ref{tab:N1N3par}.
\begin{figure}[htb!]
	\begin{center}
		\includegraphics[width=\linewidth]{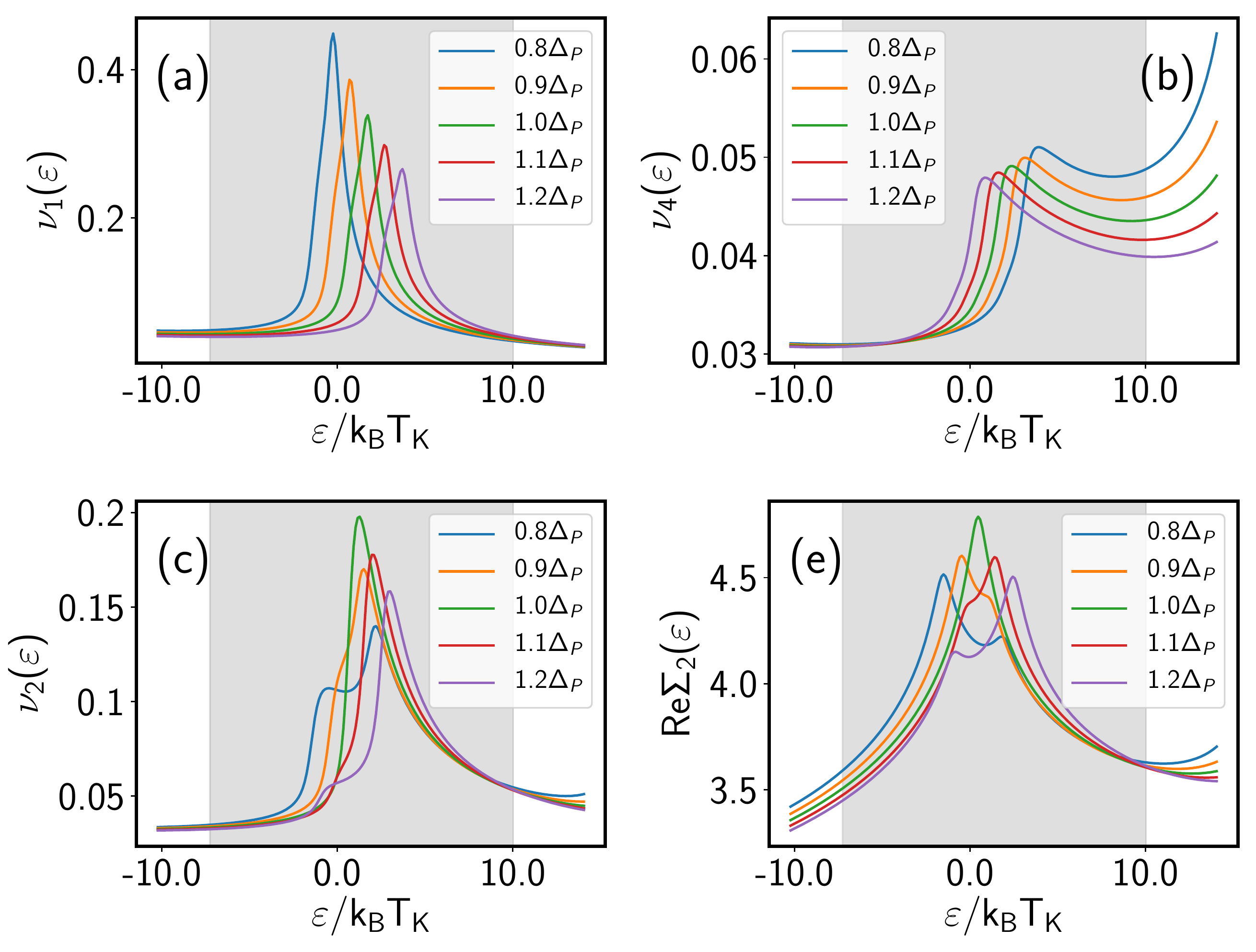}
		\caption{Channel density of states $\nu_j$, (a)-(d), and  self-energy $\Sigma_2$ of channel 2, panels (e) and (f), evaluated at bias drops $eV_{}\simeq \Delta_P$, where  $\Delta_P$ is the addition energy of the $P$-resonance, for the $N=1$ case at $B=8.05 T$. The gray stripe indicates the bias window set by the lead chemical potentials when $eV_{}=\Delta_P$. At $eV_{}=\Delta_P$ the channel density of states $\nu_2$ and $\nu_3$ are maximal and of the order of $\nu_1$. This is due  to a resonance of the associated self-energy, as illustrated in (e), (f) on the example of $\Sigma_2$. On the other hand $\nu_4$ is negligible.    } 
		\label{fig4}
	\end{center}
\end{figure}
As in the experiment, the KEA current-voltage characteristics display a $P$-peak at negative  (positive) potential drop $eV$ for valley $N=1$ ($N=3$). 

To understand the origin of the resonance, we analyze each individual TDOS $\nu_j$. In general, Kondo resonances appear in the differential conductance when a  peak in one or more of the  $\nu_j(\varepsilon)$ enters   the bias window defined by $\mu_L-\mu_R=eV_{}$. As seen from Eq. (\ref{S_e2c}), the $\nu_j$ explicitly and significantly depend on the applied bias voltage through their self-energies $\Sigma_j$. Further, peaks in $\nu_j$ originate from peaks in Re$\Sigma_j$.  At low temperatures, the latter occur when 
 $\varepsilon=\mu_\alpha+\Delta_{ji}-$Im${\cal E}$, with $i=T_j,C_j$. Simultaneously, Im$\Sigma_j$ drops by $\gamma_{{\alpha} i}$ as $\varepsilon$ is swept across the resonance. Conventional Kondo resonances, i.e., the ${\cal T}$- and $C$-resonances, arise as a consequence of a peak in $\nu_1$ entering the bias window. The mechanism for the $P$-resonance is different.  

In Fig. \ref{fig4} we focus on the resonance in the $N=1$ valley. We show the energy dependence of each $\nu_j(\varepsilon,eV_{})$ for different  potential drops  $eV\simeq \Delta (B)=\Delta_P$ for a magnetic field $B=8.05T$. The gray region indicates the bias window for asymmetric potential drop $\eta=0.4$ and $eV_{}=\Delta_P$. 
From Figs. \ref{fig4}(a) and (b), we see that $\nu_1$  is large while $\nu_4 $ is negligible in the integration window; further, $\nu_1$ exhibits a monotonic variation  as the potential drop increases. Strikingly, $\nu_2$ and $\nu_3$ develop a peak at $eV=\Delta_P$ and are of the same order of  $\nu_1$, as seen in Figs. \ref{fig4}(c) and (d). The peak reflects a resonant feature of $\Sigma_{2}$ and $\Sigma_3$, as shown in panels (e) and (f) on the example of $\Sigma_2$. This occurs because when 
$\mu_L-\mu_R=\Delta_P (=\Delta_C+\Delta_T$)  
the resonances of $\Sigma_{2}$ ($\Sigma_{3}$) at $\varepsilon=\mu_R+\Delta_T$ and $\varepsilon=\mu_L -\Delta_C$ ($\varepsilon=\mu_R+\Delta_C$ and $\varepsilon=\mu_L-\Delta_T$) merge into a single concerted resonance.
 Correspondingly, the differential magnetoconductance displays a small resonance feature also at voltages matching the condition $e V_{} =\Delta_P (B)$, as seen in Fig. \ref{fig3}. 
 While the existence of this effect is independent of asymmetries, its magnitude does depend on them. Numerically, for $N=1$ ($N=3$) we find coupling asymmetry  thresholds $\zeta_{1}$ ($\zeta_3$)  above which the resonance is seen. E.g. for the valley $N=1$ and $V<0$ it should hold that $\gamma_{L\rm{u}}-\gamma_{L\rm{d}}>\zeta_1$.  
 If the coupling strengths are reverted, $\gamma_{L\rm{d}}-\gamma_{L\rm{u}}>\zeta_1$, the resonance occurs at positive rather than at negative bias. Finally, the conditions for the single hole case, $N=3$, can be obtained from the $N=1$ case by replacing ${\rm u} \leftrightarrow \rm{d}$ and $\zeta_1$ with $\zeta_3$. Thus, if a $P$-resonance is observed at positive bias in the $N=1$ valley, it is likely that such resonance also occurs at negative bias in the $N=3$ valley, in agreement with the experimental observations. 
 Further, bias asymmetry may make it easier to observe a $P$-resonance for small values of $\mu_0-\varepsilon_d$. The reason for this is that the tails from the charge-transfer peaks  may assist the $P$-peaks if they are located at the same bias polarity. Since the asymmetry parameter $\eta$ defines the bias window for the integration variable $\varepsilon\in[\mu_0+\eta eV, \mu_0-(1-\eta)eV]$,
  $P$-peaks obtained at $\mu_L-\mu_R>0$  are assisted by charge-transfer tails for $\eta<0.5$, as seen  in Fig.~\ref{fig1} and Fig. \ref{fig2}. 

\begin{table}
\caption{Parameters used to fit the experimental data}
\begin{ruledtabular}
\begin{tabular}{lccccc}
$N$& $\Gamma$ & $\mu_0-\varepsilon_d$ & $\gamma_{L \rm d}$&$\gamma_{L \rm{u}}$ & $\eta$\\
\hline
1 & 3.5 meV & 4.83 & 0.42  & 0.08 & 0.42  \\
3 & 2.4 meV & 5.36 & 0.3 & 0.06 & 0.1 
\\
\end{tabular} \label{tab:N1N3par}
\end{ruledtabular}
\end{table}
 Note that virtual processes which involve a direct inelastic transition from  the $j=1$ to the $j=4$ channel, namely a $P$-transition, are not explicitly appearing in the KEA self-energies Eq. (\ref{S_e2c}). Pseudospin conserving cotunneling transitions become important at high temperatures and are the dominant contribution to $P$-like resonances outside the Kondo regime   \cite{Niklas2016,Jespersen2011}. 
 
\textit{Conclusion.-} 
We have observed the emergence of inelastic resonances at bias voltages corresponding to  pseudospin conserving $P$-transitions in a Kondo correlated CNT-QD. Due to the antiferromagnetic character of Kondo correlations, which inhibit direct $P$-transitions, these  resonances emerge non trivially from a coherent addition of pseudospin non-conserving  ${\cal T}$- and $C$- transitions. The here established mechanism for $P$-like resonance becomes prominent in the presence of asymmetries in the tunneling coupling and bias drop. 

A.L. and T.H. equally contributed to this work. We acknowledge support by the Deutsche Forschungsgemeinschaft within  SFB 689, SFB 1277 B04, and by JSPS KAKENHI Grant Numbers JP15J01518,
JP19H00656, and JP19H05826. 
\bibliographystyle{apsrev}
\bibliography{bibl}
\end{document}


\title{Supplementary Material: Unhiding a concealed resonance by multiple Kondo transitions in a quantum dot} 
\author{Aritra Lahiri}
\affiliation{Institut f\"ur Theoretische Physik, Universit\"at Regensburg, D-93040 Regensburg, Germany}
\affiliation{School of Physics and Astronomy, University of Minnesota, Minneapolis, MN 55455, USA}
\author{Tokuro Hata}
\affiliation{Department of Physics, Graduate School of Science, Osaka University, 560-0043 Osaka, Japan}
\author{Sergey Smirnov}
\affiliation{P. N. Lebedev Physical Institute of the Russian Academy of Sciences, 119991, Moscow, Russia}
\author{\\Meydi Ferrier}
\affiliation{Department of Physics, Graduate School of Science, Osaka University, 560-0043 Osaka, Japan}
\affiliation{Laboratoire de Physique des Solides, CNRS, Univ. Paris-Sud, Universit\'e Paris Saclay, 91405 Orsay Cedex, France}
\author{Tomonori Arakawa}
\affiliation{Department of Physics, Graduate School of Science, Osaka University, 560-0043 Osaka, Japan}
\author{Michael Niklas} 
\affiliation{Institut f\"ur Theoretische Physik, Universit\"at Regensburg, D-93040 Regensburg, Germany}
\author{Magdalena Marganska} 
\affiliation{Institut f\"ur Theoretische Physik, Universit\"at Regensburg, D-93040 Regensburg, Germany}
\author{Kensuke Kobayashi} 
\email{kensuke@meso.phys.sci.osaka-u.ac.jp}
\affiliation{Department of Physics, Graduate School of Science, Osaka University, 560-0043 Osaka, Japan}
\affiliation{Center for Spintronics Research Network (CSRN), Graduate School of Engineering Science, Osaka University, Osaka 560-8531, Japan} 
\affiliation{Institute for Physics of Intelligence and Department of Physics, The University of Tokyo, Tokyo 113-0033, Japan.
}
\author{Milena Grifoni} 
\email{milena.grifoni@physik.uni-regensburg.de}
\affiliation{Institut f\"ur Theoretische Physik, Universit\"at Regensburg, D-93040 Regensburg, Germany}
\date{\today}

\maketitle

\tableofcontents

\section{Model Hamiltonian} 
We describe a carbon nanotube (CNT) quantum dot in terms of a four-levels Anderson model, where the tunneling is diagonal in the Kramers basis, i.e. in the eigenbasis of the CNT. Then the Hamiltonian for the total leads-plus-CNT system reads 
\begin{equation}
\hat{H}_{\rm tot}=\sum_{i} \varepsilon_i(B)\hat{n_i}+U\sum_{i\neq j}\hat{n_i}\hat{n_j} +\sum_{\alpha,k,i}t_{\alpha i} \hat c^\dagger_{\alpha ik}\hat d_i +h.c. +\hat{H}_{\rm leads} , \quad i,j=1,\cdots, 4 
\label{Htot}
\end{equation}
where $d_i/d^\dagger_i$ denote the dot electron operators for the $i^{th}$ state, $\hat{n_i}=d^\dagger_i d_i$ is the occupation of the $i^{th}$ state, $U$  the charging energy.
In the case of inter-Kramers splitting larger than the Kondo temperature, $\Delta > k_{\rm B}T_K$, of relevance for the experiment, the distinct contribution from the two Kramers pairs must be considered.
To properly describe this regime we set the single particle energies such that at zero magnetic field is  
$\varepsilon_{1,2}=\varepsilon_d$ for the lower Kramers pair and $\varepsilon_{3,4}=\varepsilon_u=\varepsilon_d+\Delta$ for the upper pair.
 Further,  $c_{\alpha ik}/c^\dagger_{\alpha ik}$ denote the electron operators for the lead $\alpha$ characterized by the longitudinal wavevector $k$, and the index $i$. Finally,  $t_{\alpha i }$ is the tunneling amplitude for tunneling from lead $\alpha$ in the $i^{th}$ state. The resulting tunneling couplings are $\Gamma_{{\alpha}i}= 2\pi\vert t_{\alpha i}\vert^2 D(\varepsilon_F) $. Here $D(\varepsilon_F)$ is the lead density of states at the Fermi energy. For later convenience we introduce the 
 normalized tunneling couplings  $\gamma_{{\alpha} i}=\Gamma_{{\alpha}i}/\sum_\alpha \Gamma_{{\alpha} i}$.
 
We are interested in the Kondo effect in valley $N=1$ and $N=3$ of a CNT, corresponding to occupation with one electron or one hole, respectively. In this case it is sufficient to consider the limit of strong/infinite electron-electron interaction to capture the essential physics governing the Kondo effect in our system. In this limit, only virtual transitions to the empty dot state for valley $N=1$, or the fully filled shell for valley $N=3$, are  included.

\section{Tunneling density of states within the Keldysh effective action theory}
 
 The current through our four-levels Anderson model is conveniently obtained from the tunneling density of states (TDOS) $\nu_j=-\frac{1}{\pi}\mathrm{Im}G_j$ of level $j$ according to the Meir and Wingreen formula \cite{Meir1992,Meir1993,PhysRevB.49.11040}, see Eq. \eqref{current2} of the main text. Here $G_j(\varepsilon)$ is the Fourier transform of the retarded single particle Green's function $G_j(t)=-(i/\hbar)\theta(t)\langle \{c_{i}(t),c^\dagger_{i}\}\rangle$.
 The differential conductance $G_{\mathrm{diff}}=\frac{d}{dV}I$ is in turn given by
 \begin{align}
 G_{\mathrm{diff}}&=\frac{e^2}{\hbar}\frac{d}{d(eV)}\sum_{j=1}^{4} \int_{-\infty}^{\infty}d\varepsilon \gamma_{Lj}\gamma_{Rj}\Gamma_j\nu_j(\varepsilon)[f_L(\varepsilon)-f_R(\varepsilon)]. \label{diffgtaylor}
 \end{align}
In this work the channel TDOS $\nu_j$ is evaluated along the lines of Ref. \citen{Schmid2015} using the method of the Keldysh effective action (KEA). Within the KEA, first an infinite-$U$ slave boson transformation of the total Hamiltonian Eq. (\ref{Htot}) is performed. Then a functional field integral formulation is adopted for the problem (see Ref.~\citen{kamenev_2011,altland_simons_2006} for details). The functional integral approach is convenient as it enables one to integrate out the fermionic degrees of reservoirs and dot exactly, thus leaving an effective action which only depends on the bosonic field associated to the slave-boson operator. An expansion of the tunneling action about a non-zero slave-boson field configuration enables one to obtain an analytic expression for the TDOS.
We find, 
\begin{equation}
\nu_j(\varepsilon,\vec{B})=\frac{1}{2\pi}
\frac{\Gamma  }{[\varepsilon_j(\vec{B})-\varepsilon+\Gamma_j\text{Re}\Sigma_j(\varepsilon,\vec{B})]^2+[\Gamma_j\text{Im}\Sigma_j(\varepsilon,\vec{B})]^2},\label{TDOS}
\end{equation}
 where we explicitly
indicated that $\nu_j$ may depend on an external magnetic field $\vec{B}$. Here   $\Gamma_j=\sum_\alpha\Gamma_{\alpha j}$ and $\Gamma=(1/4)\sum_j\Gamma_j$ is the average rate. 
Furthermore, $\text{Re}\Sigma_j(\varepsilon,\vec{B})$ and $\text{Im}\Sigma_j(\varepsilon,\vec{B})$ are, respectively, the real and
imaginary parts of the self-energy,
\begin{equation}
\Sigma_j(\varepsilon,\vec{B})=-\sum_\alpha\sum_{i=1}^4 \gamma_{\alpha  i}\frac{\Gamma_i}{\Gamma_j}\int_{-\infty}^\infty\frac{d\varepsilon'}{2\pi}
\frac{ L_W(\varepsilon') f_{\alpha}(\varepsilon')}{\varepsilon'-\varepsilon + \Delta_{ji}(\vec{B})+{i}E^j_i},
\label{S_e1}
\end{equation}
with  $\Delta_{ji}=\varepsilon_j-\varepsilon_i$ the energy  difference between the eigenstates $j$ and $i$. 
Here $L_W(\varepsilon) =\frac{W^2}{\varepsilon^2+W^2}$, with $W$ being the lead bandwidth, cures  ultraviolet divergences. Finally, the functions $E_j^i$  are the product of the expansion points of the classical and quantum components of the slave-boson fields \cite{Smirnov2013}, and are a crucial ingredient of the (KEA) theory to describe the non-perturbative unitary limit of the Kondo effect. Each of the $E_j^i$ is not calculated {\em  a priori}; instead, it is fixed by imposing {\em a posteriori} constraints on the channel TDOS.    These are i) time-reversal and particle-hole conjugation relations which relate the TDOS component $\nu_j(\varepsilon,\vec{B})$ to its time-reversal and particle-hole related components; ii) the fulfillment of the Friedel sum rule \cite{Langreth1966}  with its implications.  
Such relations will be explicitly discussed in the next two sections.
\section{Self-energies for perpendicular magnetic field}
In  magnetic field the single particle energy levels $\varepsilon_j$ acquire a magnetic field dependence.  The splitting  $\Delta (B)=\varepsilon_{4}(B)-\varepsilon_1(B):=\Delta_P$ is given explicitly in Eq. \eqref{deltaB} of the main text for the case of a perpendicular magnetic field of magnitude $B$. Then one imposes the particle-hole and time-reversal conjugation relations 
\begin{eqnarray}
\nu_{1,2}(\varepsilon +\Delta(B)/2 -\varepsilon_M)&=&\nu_{4,3}(\varepsilon -(\Delta(B)/2-\varepsilon_M)), \nonumber\\ \nu_{2,3}(\varepsilon,B)&=&\nu_{1,4}(\varepsilon,-B),  
\end{eqnarray}
where $\varepsilon_M=\sum_j\varepsilon_j/4$ is the middle of the shell.  This fixes some relations among the parameters $E^j_i$ (cf. Eq. (B23) in Ref.  \citen{Schmid2015}), and leaves still just a parameter ${\cal E}$ to be determined. Since in the Keldysh field integral the slave-bosonic fields $b(t)$ and $\bar{b}(t)$ are not complex conjugate of each other, the parameter  ${\cal E}$ must be complex. From the above equations also the chiral conjugation relations follow:
\begin{eqnarray}
\nu_{1,2}(\varepsilon + \Delta(B)/2-\varepsilon_M)&=&\nu_{3,4}(\varepsilon -(\Delta(-B)/2-\varepsilon_M)).
\end{eqnarray}
 Using such parametrization the  equation for the self-energies Eq. \eqref{S_e1} turns into
\begin{equation}
\Sigma_{j }(\varepsilon,B)=\sum_\alpha \sum_{ i=T_j, C_j} 2\gamma_{\alpha i} \frac{\Gamma_i}{\Gamma_j}K_\alpha (\varepsilon,\Delta_{ji }),
\label{S_e2b}
\end{equation}
where we introduced the notation  $T_j$, $C_j$ for the states associated to the state $j$ by time-reversal and chiral conjugation, respectively. Further, 
\begin{equation}
K_{\alpha} (\varepsilon,\Delta ) =-\int_{-\infty}^\infty\frac{d\varepsilon'}{2\pi}
\frac{f_{\alpha}(\varepsilon')L_W(\varepsilon')}{\varepsilon'-\varepsilon +\Delta +{i}{\cal E}}
= \frac{i}{4} + \frac{1}{2\pi}\Bigg[ {\rm Re}\Psi \left( \frac{1}{2} + iz_\alpha \right)-\Psi\left( \frac{1}{2} +\frac{\cal E}{2\pi k_{\rm B}T} -i\frac{\mu_\alpha -\varepsilon +\Delta}{2\pi k_{\rm B}T}   \right)\Bigg], 
\label{kappa}
\end{equation}
where $\Psi (x) $ is the digamma function and $iz_\alpha=- (i \mu_\alpha +W)/2\pi k_{\rm B} T$. Also, $\mu_\alpha=\mu_0 +eV_\alpha$, with $V_L-V_R=V$ the applied bias, are the lead electrochemical potentials. 
The complex function ${\cal E}$ is now fully determined by the remaining requirements (set by the Friedel sum rule) to be fulfilled by the TDOS;  
the resulting equations for $\cal{E}$ are explicitly derived in the next section.

In the limit of large cut-offs $W$ the first digamma function in  Eq. \eqref{kappa} simplifies to Re$\Psi(1/2 +iz_\alpha)\approx \ln (W/2\pi k_{\rm B}T)$, which is independent of the lead index $\alpha$. From Eq. \eqref{S_e2b} we thus find  for the self-energies the form,
\begin{equation}
\Sigma_{j }(\varepsilon,B)=\frac{1}{\pi } \sum_{ i=T_j, C_j}\frac{\Gamma_i}{\Gamma_j}\bigg[ \ln \left( \frac{W}{2\pi k_{\rm B}T} \right) + \frac{i \pi}{2}- \sum_\alpha \gamma_{\alpha i}\Psi\left( \frac{1}{2} +\frac{\cal E}{2\pi k_{\rm B}T} -i\frac{\mu_\alpha -\varepsilon +\Delta_{ji}}{2\pi k_{\rm B}T}   \right)\bigg].\label{Selfeform}
\end{equation}
This is Eq. \eqref{S_e2c} of the main text. 
We then find
\begin{align}
\label{reimfiniteeps}
\Gamma_j{\rm{Re}}\Sigma_j(\varepsilon)&=\frac{1}{\pi } 
\sum_{ i=T_j, C_j}\Gamma_i\bigg[ \ln \left(
 \frac{W}{2\pi k_{\rm B}T} \right) - \sum_\alpha \gamma_{\alpha i} 
 {\rm{Re}}
\Psi\left( \frac{1}{2} +\frac{\cal E}{2\pi k_{\rm B}T} -i\frac{\mu_\alpha -\varepsilon +\Delta_{ji}}{2\pi k_{\rm B}T}   
\right)\bigg],
\nonumber\\
\Gamma_j{\rm{Im}}\Sigma_j(\varepsilon)&=\frac{1}{\pi } \sum_{ i=T_j, C_j}\Gamma_i\bigg[\frac{\pi}{2}- \sum_\alpha \gamma_{\alpha i}{\rm{Im}}
\Psi\left( \frac{1}{2} +\frac{\cal E}{2\pi k_{\rm B}T} -i\frac{\mu_\alpha -\varepsilon +\Delta_{ji}}{2\pi k_{\rm B}T}   \right)
\bigg].
\end{align}

\section{Unitary conductance conditions}
In this section we derive the equations that $\cal E$ has to satisfy in the SU(4) and SU(2)$\otimes$SU(2) cases, reached for $k_{\rm B}T_K\gg \Delta$ and $k_{\rm B}T_K \leqslant \Delta$, respectively.  
Few relations involving the digamma functions listed below are required for the derivation.
\begin{align}
\Psi\left(\frac{1}{2}+\frac{{ \cal E }}{2\pi k_{\rm B}T} \right)\bigg\rvert_{T\to 0}&\approx \mathrm{ln}\left(\frac{ \cal E  }{2\pi k_{\rm B} T }\right)+\frac{1}{24}\left(\frac{2\pi k_{\rm B} T}{\cal E}\right)^2\label{psilargezlim}\\
\implies T\frac{\partial }{\partial T}\Psi\left(\frac{1}{2}+\frac{\cal E}{2\pi k_{\rm B} T} \right)\bigg\rvert_{T=0}&=T\left(-\frac{d}{dT}\mathrm{ln}T+\frac{1}{24}\left(\frac{2T(2\pi k_{\rm B} )}{\cal E}\right)^2\right)\bigg\rvert_{T=0}=-1 \,.\label{ddTpsi}
\end{align}
Further, the following chain relations hold 
\begin{align}
\frac{\partial}{\partial T}\Psi\left(\frac{1}{2}+\frac{\cal E}{2\pi k_{\rm B} T}+\frac{i\varepsilon}{2\pi k_{\rm B}T} \right)&= \Psi\left(\frac{1}{2}+\frac{\cal E}{2\pi k_{\rm B} T}+\frac{i\varepsilon}{2\pi k_{\rm B}T} \right)'\left(-\frac{{\cal E} +i\varepsilon}{2\pi k_{\rm B} T^2}\right) \,, \label{deldelTpsi}\\
\frac{\partial}{\partial\varepsilon}\Psi\left(\frac{1}{2}+\frac{\cal E}{2\pi k_{\rm B} T }+\frac{i\varepsilon}{2\pi k_{\rm B}T} \right)&= \Psi\left(\frac{1}{2}+\frac{\cal E}{2\pi k_{\rm B} T}+\frac{i\varepsilon}{2\pi k_{\rm B}T} \right)'\frac{i}{2\pi k_{\rm B}T}\nonumber\\
&=-\frac{i}{{\cal E}+i\varepsilon}T\frac{\partial}{\partial T}\Psi\left(\frac{1}{2}+\frac{\cal E}{2\pi k_{\rm B} T}+\frac{i\varepsilon}{2\pi k_{\rm B}T} \right) \,.\label{deldelepsi0}
\end{align}
Therefore, using Eqns.~\eqref{deldelTpsi} and~\eqref{deldelepsi0},
\begin{align}
\frac{\partial}{\partial\varepsilon}\Psi\left(\frac{1}{2}+\frac{\cal E}{2\pi k_{\rm B} T}+\frac{i(\varepsilon -\mu_0)}{2\pi k_{\rm B}T} \right)\bigg\rvert_{\substack{\varepsilon=\mu_0 \\T=0}}
&=\frac{i}{\cal E}
= \frac{\mathrm{sin}\varphi+i\mathrm{cos}\varphi}{|{\cal E}|}\,,\label{deldelepsi}
\end{align}
where we expressed ${\cal E}=\vert {\cal E}\vert e^{i\varphi}$. Similarly, 
\begin{align}
\frac{\partial^2}{\partial\varepsilon^2}\Psi\left(\frac{1}{2}+\frac{\cal E}{2\pi k_{\rm B} T}+\frac{i(\varepsilon -\mu_0)}{2\pi k_{\rm B}T} \right)\bigg\rvert_{\substack{\varepsilon=\mu_0 \\T=0}}&=\frac{\mathrm{cos}(2\varphi)-i\mathrm{sin}(2\varphi)}{{|\cal E|}^2}\,.\label{del2dele2psi}
\end{align}
We are now in the position of fixing  the real and imaginary parts of ${\cal E}$. To this extent 
we use the Friedel sum rule \cite{Langreth1966}. For the case of symmetrically coupled leads,  $\gamma_{\alpha i}=1/2$, it reads
	\begin{equation}
	{\cal T}_j(\mu_0)=\frac{\pi}{2}\Gamma_j\nu_j(\varepsilon_0)=\sin^2 \delta_j. \label{trans}
	\end{equation}  It relates the transmission  per channel at the Fermi level ${\cal T}_j(\mu_0)$ to the scattering phase shift $\delta_j=\pi\langle n_j\rangle$, with $\langle n_j \rangle$ the average occupation of the level $j$. This in turn determines the zero temperature conductance through
	$G_0=(e^2/h)\sum_j	{\cal T}_j(\mu_0) $. 	
	The mean occupation depends on the energy degeneracy of a given channel. For example, in the SU(4) case with four-fold degeneracy and in the valley with $N=1$ is $\langle n_j \rangle =1/4 $ for each $j=1,..,4$. As a consequence, ${\cal T}_j(\mu_0)=1/2$ and $G_0=2e^2/h$ \cite{Choi2005}. Because the maximum value the transmission can take is one, this also means that the maximum of the transmission for the SU(4) does not lie at the Fermi level, but is actually shifted from it by a shift $\kappa \approx k_{\rm B}T_K$. This shift is positive for $N=1$ and negative for $N=3$ \cite{hewson_1993}. 
	
	For the SU(2)$\otimes$SU(2) case 
	one needs the occupations of the lower and upper Kramers doublets  $\langle n_{1,2} \rangle :=\langle n_{d} \rangle $ and  $\langle n_{3,4} \rangle :=\langle n_{u} \rangle $, respectively. This yields the phase shifts 
	$\delta_{u,d}=\pi \langle n_{u,d} \rangle$.
	When $\Delta\gg k_{\rm B}T_K$ at zero temperature only the lower Kramers doublet will be occupied, 
	 $\langle n_{d} \rangle = 1/2 $, with the upper being empty, $\langle n_{3,4} \rangle :=\langle n_{u} \rangle = 0 $.
	 This yields ${\cal T}_{1,2}(\mu_0)=1$,  ${\cal T}_{3,4}(\mu_0)=0$ and $G_0=2e^2/h$. As a consequence, matching the SU(2) behavior, the transmission is dominated by the lower Kramers doublet and is maximal at the Fermi level. For the intermediate case with  $\Delta\approx k_{\rm B}T_K$ the situation is more complex since both Kramers doublets are occupied. For generic $\Delta$ one can express the unbalance in the occupation between the lower and upper Kramers doublet in terms of the Kramers pseudospin magnetization $\delta n$ defined by   
	  $ \langle n_d \rangle  = 1/4+ \delta n $ and $\langle n_{u} \rangle = 1/4 - \delta n $.  Importantly, it still holds   $\sin^2\delta_d +\sin^2\delta_u=1$,  
	  which  ensures $G_0=2e^2/h$ independent of the value of $\Delta$ \cite{Sakano2006,Mantelli2016}. On the other hand  a finite polarization fixes the difference in the transmission at the Fermi level through $\sin^2\delta_d-\sin^2\delta_u=\sin(2\pi \delta n)$. The zero temperature equilibrium occupations $\langle n_{u,d}\rangle$ and hence $\delta n$ can be evaluated exactly for generic $\Delta$ through the Bethe-Ansatz method \cite{Sakano2006}. We postpone  using proper Bethe-Ansatz equations for the determination of $\langle n_{u,d}\rangle$ to future analysis. Instead, here  an approximation which accounts for the shifts and is appropriate for the parameter regime $\Delta > k_B T_K$ of our experiment will be discussed. \\  
According to the above considerations, 
\begin{itemize}
	\item
First we impose that in the limit of zero temperature, zero bias, zero magnetic field and for symmetric couplings the differential conductance of the CNT quantum dot  reaches the correct unitary limit $G_0=2e^2/h$.  The differential conductance $G_{\mathrm{diff}}=\frac{d}{dV}I$ is given by
\eqref{diffgtaylor}. 
In the symmetric case is $\gamma_{{\alpha} j}=1/2$. It thus must hold  
\begin{align}
G_0=\lim_{T,V,B \to 0}G_{\mathrm{diff}}&=\frac{e^2}{\hbar}               \sum_j\frac{\Gamma_j}{4}\nu_j(\varepsilon) \Bigg \rvert_{\varepsilon=\mu_0} 
= \frac{2e^2}{h} 
\implies\ \frac{\pi }{2}(\Gamma_d \nu_d(\mu_0)+\Gamma_u\nu_u(\mu_0)) =1, \label{TDOS cond1}
\end{align}
where we set $\nu_{1,2}=\nu_d$ and $\nu_{3,4}=\nu_u$ for the TDOS of the lower and upper Kramers doublet, respectively.

To evaluate $\nu_j(\mu_0)$ we need the self-energies in this limit.  
From \eqref{Selfeform} we get 
\begin{equation}
\Sigma_{j }(\mu_0)=\frac{1}{\pi } \sum_{ i=T_j, C_j}\frac{\Gamma_i}{\Gamma_j}\bigg[ \ln \left( \frac{W}{2\pi k_{\rm B}T} \right) + \frac{i \pi}{2}- \Psi\left( \frac{1}{2} +\frac{\cal E}{2\pi k_{\rm B}T} -i\frac{\Delta_{ji}}{2\pi k_{\rm B}T}   \right)\bigg],\label{Selfeform0}
\end{equation}
with contributions only from the chiral ($C_j$) and time-reversed ($T_j$) partners. The zero temperature channel TDOS are hence given by,
\begin{align}
\nu_d(\mu_0)&=\frac{\pi}{8\Gamma_d^2}\frac{\Gamma}{\left[\frac{(\varepsilon_d-\mu_0)\pi}{2\Gamma_d} + \frac{1}{2}\frac{\Gamma_u}{\Gamma_d}\mathrm{ln}\left(\frac{W}{\vert{\cal E }+ i\Delta\vert}\right)
	+\frac{1}{2}\mathrm{ln}\left(\frac{W}{\mathopen|{\cal E}\mathclose|}\right) \right]^2+\left[ \frac{\pi}{4}\left(1+\frac{\Gamma_u}{\Gamma_d}\right)-\frac{1}{2}\left(\varphi+\frac{\Gamma_u}{\Gamma_d}{\varphi_{+}}\right)\right]^2}\,, \label{tdoslevelscheme1}
\end{align}
where $\varphi=\rm{Arg}\{{\cal E}\}$, and $\varphi_{+}=\rm{Arg}\{{\cal E} +i\Delta \}$. Similarly,
\begin{align}
\nu_u(\mu_0)&=\frac{\pi}{8\Gamma_u^2}\frac{\Gamma}{\left[\frac{(\varepsilon_u-\mu_0)\pi}{2\Gamma_u} + \frac{1}{2}\frac{\Gamma_d}{\Gamma_u}\mathrm{ln}\left(\frac{W}{\vert{\cal E }- i\Delta\vert}\right)
	+\frac{1}{2}\mathrm{ln}\left(\frac{W}{\mathopen|{\cal E}\mathclose|}\right) \right]^2+\left[ \frac{\pi}{4}\left(1+\frac{\Gamma_d}{\Gamma_u}\right)-\frac{1}{2}\left(\varphi+\frac{\Gamma_d}{\Gamma_u}{\varphi_{-}}\right)\right]^2}\,, \label{tdoslevelschemeup}
\end{align}
with $\varphi_{-}=\rm{Arg}\{{\cal E} -i\Delta \}$.
\\

%
\item Second, we locate the peaks of the zero temperature and zero bias channel TDOS  for the lower/upper Kramers channels appropriately. To this extent we introduce a parameter $\delta_j$ such that  $\delta_j=\kappa_d$ if $j\in d$ and
$\delta_j = \Delta+\kappa_u$ if $j\in u$. Further, for simplicity we approximate $\kappa_d\approx \kappa_u=\kappa$, and impose the condition

\begin{equation}
\frac{d\nu_d}{d\varepsilon} \Bigg \rvert_{\substack{\varepsilon=\mu_0+\delta_d\\
V=T=0}} =0,\qquad \frac{d\nu_u}{d\varepsilon} \Bigg \rvert_{\substack{\varepsilon=\mu_0+\delta_u\\
V=T=0}} =0.  \label{tdoslevelschemeup2}
\end{equation}
\end{itemize}
\subsection{Constraints for the SU(4) case}
The SU(4) case is characterized by fourfold degenerate levels, $\varepsilon_j=\varepsilon_d$, and by  $\Gamma_j=\Gamma$. Hence  for vanishing inter-Kramers splitting, 
$\Delta=0$, one gets from Eq. (\ref{tdoslevelscheme1})  the simpler SU(4) form
\begin{align}
\nu_j(\mu_0)&=\frac{\pi}{8\Gamma}\frac{1}{\left[\frac{(\varepsilon_d-\mu_0)\pi}{2\Gamma} + \mathrm{ln}\left(\frac{W}{\mathopen|{\cal E}|}\right) \right]^2+\left[ \frac{\pi}{2}-\varphi_{}\right]^2}\,, \label{tdoslevelscheme1su4}
\end{align}
being independent of $j$. We introduce now the SU(4) Kondo temperature 
$k_{\rm B}T_K =2W\text{exp} \big(\pi(\varepsilon_d-\mu_0)/2\Gamma\big)$
to find
\begin{align}
\nu_j(\mu_0)&=\frac{\pi}{8\Gamma}\frac{1}{\left[ \mathrm{ln}\left(\frac{k_{\rm B}T_K}{2\mathopen|{\cal E}|}\right) \right]^2+\left[ \frac{\pi}{2}-\varphi\right]^2}\,. \label{tdoslevelscheme1su5}
\end{align}
Thus, according to Eq. (\ref{TDOS cond1}), one finds the first condition on ${\cal E}$ for the SU(4)  case
\begin{equation}
\label{su4cond1k}
\mathrm{ln}\left(\frac{2\mathopen|{\cal E}|}{2k_{\rm B}T_K}\right)^2+\left[ \frac{\pi}{2}-\varphi\right]^2=\frac{\pi^2}{8}\;.
\end{equation}
For the second condition we impose that $\nu_j(\varepsilon)$ has a peak at $\mu_0 +\kappa$ and notice that $\nu_j$ is independent of $j$ in the SU(4) case.  
 From the vanishing of the numerator $N$ of the derivative of $\nu_j$ it follows
\begin{align}
N & = 2\left[(\varepsilon_d-\varepsilon)+\Gamma\mathrm{Re}\Sigma(\varepsilon)\right] \left(-1+\Gamma\frac{d \mathrm{Re}\Sigma}{d\varepsilon} \right)\Bigg\rvert_{\substack{\varepsilon=\mu_0 +\kappa\\ V=0\\T=0}} + 2\left[ \Gamma \mathrm{Im} \Sigma\right]\left(\Gamma\frac{d{\rm Im}\Sigma(\varepsilon)}{d\varepsilon}\right)\Bigg\rvert_{\substack{\varepsilon=\mu_0 +\kappa\\ V=0\\T=0}} =0.\nonumber\\
\end{align}
Hence,  on using 
${\cal E}+i\kappa =\mathopen|{\cal E}+i\kappa \mathclose|\exp\{i\varphi(\kappa)\}$, we find 
\begin{align}
 \left[-\frac{\kappa\pi}{2\Gamma}-\mathrm{ln}\left(\frac{2\mathopen|{\cal E}+i\kappa\mathclose|}{k_{\rm B}T_K}\right)\right]\left(\frac{\pi\mathopen|{\cal E} +i\kappa\mathclose|}{\Gamma}+2\mathrm{sin}\varphi (\kappa)\right)+\left(\frac{\pi}{2}-\varphi (\kappa)\right)2\mathrm{cos}\varphi(\kappa)=0.\label{su4cond2k}
\end{align}
In summary, \eqref{su4cond1k} and~\eqref{su4cond2k} together with the SU(4) shift $\kappa\approx k_{\rm B}T_K$  determine the value ${\cal E}/k_{\rm B}T_K$. These equations are universal.  In contrast, the explicit value of $T_K$ depends on microscopic details, like the actual values of  $\mu_0-\varepsilon_d$ and $\Gamma$. \\

\subsection{Constraints for the   SU(2)$\otimes$SU(2) case}
In the case of finite inter-Kramers splitting $\Delta > k_{\rm B}T_K$, of relevance for the experiment, the distinct contribution from the two Kramers pairs must be considered.
The Kondo temperature $T_K(\Delta)$ depends on $\Delta$ and one recovers the SU(4) one for vanishing splitting and equal couplings $\Gamma_j=\Gamma$. 
Further, $T_K(\Delta)$ decreases towards its smaller  SU(2) value as $\Delta $ increases \cite{Mantelli2016}. 
In the SU(2)$\otimes$SU(2) regime the conditions on $\cal E$ are  found along the lines followed for the degenerate SU(4) case. 
On the one hand the unitary condition (\refeq{TDOS cond1}) is imposed, with $\nu_d$ and $\nu_u$ given by Eqs. (\ref{tdoslevelscheme1}) and (\ref{tdoslevelschemeup}), respectively. Further,  as discussed above,  
  a finite $\Delta$ requires in principle two distinct shifts $\kappa_u $ and $\kappa_d$ associated to the upper  and lower  Kramers pair, respectively \cite{Sakano2006}.
In the following we explicitly derive the constraints imposed by Eq. \eqref{tdoslevelschemeup2}. 
Explicitly,  
\begin{align}
&\frac{d\nu_j(\varepsilon)}{d\varepsilon}\Bigg\rvert_{\substack{\varepsilon=\mu_0+\delta_j\\ V=0\\T=0}}=\left(\frac{-\Gamma}{2\pi}\right)\frac{\frac{d}{d\varepsilon}\left\lbrace\left[\varepsilon_j-\varepsilon+\Gamma_j\mathrm{Re}\Sigma_j(\varepsilon)\right]^2+\left[\Gamma_j{\rm Im}\Sigma_j(\varepsilon)\right]^2\right\rbrace}{\left\lbrace\left[\varepsilon_j-\varepsilon+\Gamma_j\mathrm{Re}\Sigma_j(\varepsilon)\right]^2+\left[\Gamma_j{\rm Im}\Sigma_j(\varepsilon)\right]^2\right\rbrace^2}\Bigg\rvert_{\substack{\varepsilon=\mu_0+\delta_j\\ V=0\\T=0}}\nonumber\\
&=\left(\frac{-2\Gamma}{2\pi}\right)\left(\frac{\pi^2}{4\Gamma_j^2}\right)^2\frac{\left[\left(-1+\Gamma_j\frac{d\mathrm{Re}\Sigma_j(\varepsilon)}{d\varepsilon}\right)\left(\varepsilon_j-\varepsilon+\Gamma_j\mathrm{Re}\Sigma_j(\varepsilon)\right)+\left(\Gamma_j\frac{d{\rm Im}\Sigma_j(\varepsilon)}{d\varepsilon}\right)\left(\Gamma_j{\rm Im}\Sigma_j(\varepsilon)\right)\right]\Bigg\rvert_{\substack{\varepsilon=\mu_0+\delta_j\\ V=0\\T=0}}}{\left\lbrace \left[\left(\frac{\varepsilon_j-\mu_0-\delta_j}{2\Gamma_j}\right)\pi+\frac{1}{2}\mathrm{ln}\left(\frac{W}{\mathopen|{\cal E}+i\delta_j\mathclose|}\right)+\frac{1}{2}\frac{\Gamma_{C_j}}{\Gamma_j}\mathrm{ln}\left(\frac{W}{\mathopen|{\cal E}+i\delta_j- i\Delta_{j,C_j}\mathclose|}\right) \right]^2 + \left[\frac{\pi}{4}\left(1+\frac{\Gamma_{C_j}}{\Gamma_j}\right)-\frac{1}{2}\left(\varphi+
\frac{\Gamma_{C_j}}{\Gamma_j}	\frac{\varphi_{j,C_j}}{2}\right)\right]^2 \right\rbrace^2}.\nonumber\\
\end{align}
It is convenient to introduce the abbreviations
\begin{align}
a_j(\mu_0+\delta_j)&=\frac{\pi}{2\Gamma_j}\left[\varepsilon_j-\varepsilon+\Gamma_j\mathrm{Re}\Sigma_j(\varepsilon)\right]\Bigg\rvert_{\substack{\varepsilon=\mu_0+\delta_j\\ V=T=0}}=\left(\frac{\varepsilon_j-\mu_0-\delta_j}{2\Gamma_j}\right)\pi+\frac{1}{2}\mathrm{ln}\left(\frac{W}{\mathopen|{\cal E}+i\delta_j\mathclose|}\right)+\frac{1}{2}\frac{\Gamma_{C_j}}{\Gamma_j}\mathrm{ln}\left(\frac{W}{\mathopen|{\cal E}+i\delta_j- i\Delta_{j,C_j}\mathclose|}\right) ,\nonumber\\
b_j(\mu_0+\delta_j)&=\frac{\pi}{2\Gamma_j}\Gamma_j{\rm Im}\Sigma_j(\varepsilon)\Bigg\rvert_{\substack{\varepsilon=\mu_0+\delta_j\\ V=T=0}}=
\frac{\pi}{4}\left(1+\frac{\Gamma_{C_j}}{\Gamma_j}\right)-\frac{1}{2}\left(\varphi+
\frac{\Gamma_{C_j}}{\Gamma_j}	\varphi_{j,C_j}\right).
\end{align}
This yields the compact expression 
\begin{align}
\frac{d\nu_j(\varepsilon)}{d\varepsilon}\Bigg\rvert_{\substack{\varepsilon=\mu_0+\delta_j \\ V=0\\T=0}}&=-\frac{\Gamma}{\Gamma_j}\left(\frac{\pi}{2\Gamma_j}\right)^2\frac{1}{\left(a_j^2(\mu_0+\delta_j)+b_j^2(\mu_0+\delta_j)^2\right)^2}
\nonumber\\&
                      \times \left[a_j (\mu_0+\delta_j) 
\left(-1+\Gamma_j\frac{d\mathrm{Re}\Sigma_j(\varepsilon)}{d\varepsilon} \Bigg\rvert_{\substack{\varepsilon=\mu_0 +\delta_j\\ V=0\\T=0}}\right) +  b_j(\mu_0+\delta_j)\left(\Gamma_j\frac{d\mathrm{Im}\Sigma_j(\varepsilon)}{d\varepsilon}\Bigg\rvert_{\substack{\varepsilon=\mu_0+\delta_j \\ V=0\\T=0}} \right)\right].
\end{align}
The derivative of Re$\Sigma$ is found  along the same lines followed for Eq.~\eqref{deldelepsi},
\begin{align}
\frac{d\Sigma_j(\varepsilon)}{d\varepsilon}\Bigg\rvert_{\substack{\varepsilon=\mu_0+\delta_j\\ V=0\\T=0}}&=\frac{-1}{\pi}\left[\frac{d}{d\varepsilon}\Psi\left(\frac{1}{2}+
\frac{{\cal{E}}-i(\mu_0-\varepsilon)}{2\pi k_{\rm B}T}\right)+\frac{\Gamma_{C_j}}{\Gamma_j}\frac{d}{d\varepsilon}\Psi\left(\frac{1}{2}+\frac{{\cal{E}}-i(\mu_0-\varepsilon +\Delta_{j,C_j})}{2\pi k_{\rm B}T}\right) \right]\Bigg\rvert_{\substack{\varepsilon=\mu_0+\delta_j \\T=0}}\nonumber\\
&=\frac{-1}{\pi}\left[ \frac{i\mathrm{cos}(\varphi(\delta_j))+\mathrm{sin}(\varphi(\delta_j))}{\mathopen|{\cal E}+i\delta_j\mathclose|}+\frac{\Gamma_{C_j}}{\Gamma_j}\frac{i\mathrm{cos}(\varphi_{j,C_j})+\mathrm{sin}(\varphi_{j,C_j})}{\mathopen|{\cal{E}}+i\delta_j-i\Delta_{j,C_j}\mathclose|} \right].
\end{align}
Notice that here is  $\varphi_{j,C_j}=\varphi_{j,C_j}(\delta_j)$.  Consequently,
\begin{align}
\frac{d\nu_j(\varepsilon)}{d\varepsilon}\Bigg\rvert_{\substack{\varepsilon=\mu_0 +\delta_j\\ V=0\\T=0}}&=\frac{\Gamma}{\Gamma_j}\frac{\pi}{4\Gamma_j}\left(\frac{1}{a_j^2(\mu_0+\delta_j)+
	b_j^2(\mu_0+\delta_j)}\right)^2\nonumber\\
&\times\left[a_j\left(\frac{\pi}{\Gamma_j} +\frac{\mathrm{sin}(\varphi (\kappa_j))}{\mathopen|{\cal E}+i\delta_j\mathclose|}+\frac{\Gamma_{C_j}}{\Gamma_j}\frac{\mathrm{sin}(\varphi_{j,C_j})}{\mathopen|{\cal E}+i\delta_j-i\Delta_{j,C_j}\mathclose|} \right)+b_j\left( \frac{\mathrm{cos}(\varphi (\delta_j))}{\mathopen|{\cal E}+i\delta_j\mathclose|}+\frac{\Gamma_{C_j}}{\Gamma_j}\frac{\mathrm{cos}(\varphi_{j,C_j})}{\mathopen|{\cal E}+i\delta_j-i\Delta_{j,C_j}\mathclose|} \right)\right]. \label{derivTDOSsu4}
\end{align}
Finally, the peak of each TDOS is set appropriately according to Eq. (\refeq{tdoslevelschemeup2}). 
Using now $\delta_{1,2}=\delta_d=\kappa$, $\delta_{3,4}=\delta_u=\Delta+\kappa$ we get the final conditions

\begin{align}
&a_d(\mu_0+\kappa)\underbrace{\left(\frac{\pi}{\Gamma_d} +\frac{\mathrm{sin}(\varphi (\kappa))}{\mathopen|{\cal E}+i\kappa\mathclose|}+\frac{\Gamma_{u}}{\Gamma_d}\frac{\mathrm{sin}(\varphi_{+}(\kappa))}{\mathopen|{\cal E}+i\kappa+i\Delta\mathclose|} \right)}_{\lambda_d (\kappa)}
+b_d(\mu_0+\kappa)\underbrace{\left( \frac{\mathrm{cos}(\varphi (\kappa))}{\mathopen|{\cal E}+i\kappa\mathclose|}+\frac{\Gamma_{u}}{\Gamma_d}\frac{\mathrm{cos}(\varphi_{+}(\kappa))}{\mathopen|{\cal E}+i\kappa+i\Delta\mathclose|} \right)}_{\omega_d(\kappa)} =0\nonumber\\
&a_u(\mu_0+\Delta+\kappa)\underbrace{\left(\frac{\pi}{\Gamma_u} +\frac{\mathrm{sin}(\varphi_+ (\kappa))}{\mathopen|{\cal E}+i\Delta+i\kappa\mathclose|}+\frac{\Gamma_{d}}{\Gamma_u}\frac{\mathrm{sin}(\varphi(\kappa))}{\mathopen|{\cal E}+i\kappa\mathclose|} \right)}_{\lambda_u (\kappa)}
+b_u(\mu_0+\Delta+\kappa)\underbrace{\left( \frac{\mathrm{cos}(\varphi_+ (\kappa))}{\mathopen|{\cal E}+i\Delta+i\kappa\mathclose|}+\frac{\Gamma_{d}}{\Gamma_u}\frac{\mathrm{cos}(\varphi(\kappa))}{\mathopen|{\cal E}+i\kappa\mathclose|} \right)}_{\omega_u(\kappa_u)}=0. \label{derivTDOSsu2}
\end{align}
Here we used $\Delta_{j,C_j}=\pm\Delta$ for $j\in u/d$, the upper/lower Kramers pair, and we introduced 
 $\varphi_{+}(\kappa)={\rm Arg}({\cal E}+ i\Delta+i\kappa)$.
The three equations \eqref{TDOS cond1} and the pair given by \eqref{derivTDOSsu2} are solved to yield Re${\cal E}$, Im${\cal E}$ and $\kappa$. 

\section{Low bias expansion of the differential conductance}
To analyze the effect of the asymmetries on the low-bias behavior, a Taylor expansion of the differential conductance in the applied bias difference $eV$ is sought: 
\begin{align}
G_{\mathrm{diff}}(eV)&=G_0+ G_1(eV)  +\ldots \label{diffgtaylorform}
\end{align}
where $G_{\mathrm{diff}}=e\frac{d}{d(eV)}I$ is given by Eq. \eqref{diffgtaylor}.
The presence of a term linear in the applied bias signifies an asymmetric response.
 Along the lines of the previous section, we introduce the coefficients
 \begin{align}
a_j(\varepsilon,T,eV) &=\frac{\pi(\varepsilon_j-\varepsilon)}{2\Gamma_j}+
\frac{1}{2}\sum_{i=C_j,T_j}\frac{\Gamma_i}{\Gamma_j}
\left[
\mathrm{ln}\left(\frac{W}{2\pi k_BT}\right)-\sum_{\alpha=L,R}\gamma_{\alpha i}\mathrm{Re}\Psi\left(\frac{1}{2}+\frac{\cal E}{2\pi k_BT}-\frac{i(\mu_\alpha-\varepsilon+\Delta_{ji})}{2\pi k_{\rm B}T} \right)\right] , \label{genb1xsu2su2}\\
b_j(\varepsilon,T,eV)&=\frac{1}{2}\sum_{i=C_j,T_j}\frac{\Gamma_i}{\Gamma_j}
\left[\frac{\pi}{2}-\sum_{\alpha=L,R}\sum_{i=C_j,T_j}\gamma_{\alpha i}\mathrm{Im}\Psi\left(\frac{1}{2}+\frac{\cal E}{2\pi k_{\rm B}T}-\frac{i(\mu_\alpha-\varepsilon+\Delta_{ji})}{2\pi k_{\rm B}T} \right)\right], \label{genb2xsu2su2}
\end{align}
which permits one to write the channel TDOS in the compact form 
\begin{align}
\nu_j(\varepsilon)&=
\frac{\Gamma}{\Gamma_j}\frac{\pi}{8\Gamma_j}\frac{1}{a_j^2(\varepsilon)+b_j^2(\varepsilon)}\label{tdosb1b2su2su2}.
\end{align}
Introducing 
$g(\varepsilon)=\sum_j\frac{\Gamma}{\Gamma_j}\frac{\gamma_{Lj}\gamma_{Rj}}{a_j^2(\varepsilon)+b^2_j(\varepsilon)}$, we get the convenient expression
\begin{align}
G_{\mathrm{diff}}&=\frac{e^2}{\hbar}\frac{\pi}{8}\frac{d}{d(eV)}\int_{-\infty}^{\infty}d\varepsilon g(\varepsilon)  [f_L(\varepsilon)-f_R(\varepsilon)].\label{diffgtaylorsu2su2sn}
\end{align}
The low bias expansion of the zero temperature differential conductance is thus,
\begin{align}
G_{\mathrm{diff}}&=\frac{e^2}{\hbar}
\frac{\pi}{8}\frac{d}{d(eV)}\int_{\mu_R}^{\mu_L}d\varepsilon g(\varepsilon,T,eV) \nonumber\\
&=\frac{e^2 }{\hbar}\frac{\pi}{8}\Bigg[\frac{d}{d(eV)}\int_{\mu_R}^{\mu_L}d\varepsilon g(\varepsilon,T,eV)\Bigg\rvert_{\substack{V=0\\T=0}}+\frac{d^2}{d(eV)^2}\int_{\mu_R}^{\mu_L}d\varepsilon g(\varepsilon,T,eV)\Bigg\rvert_{\substack{V=0\\T=0}}(eV)+\ldots\Bigg],\label{diffgtaylorsu2su2s}
\end{align}
where  the temperature and bias dependence of the function $g$ has been expressed explicitly. 
\subsection{The linear conductance $G_0$ }

Recalling $\mu_L=\mu_0+\eta eV$ and $\mu_R=\mu_0-(1-\eta)eV$ it follows from Eq. \eqref{diffgtaylorsu2su2s}
\begin{align}
\frac{d}{d(eV)}\int_{\mu_0-(1-\eta)eV}^{\mu_0+\eta eV}d\varepsilon g(\varepsilon,T,eV)\Bigg\rvert_{\substack{V =0\\T=0}}&=\left[g(\mu_0+\eta eV,T,eV)\eta + g(\mu_0-(1-\eta)eV,T,eV)(1-\eta) \right]\Bigg\rvert_{\substack{V =0\\T=0}}\nonumber\\
&+\int_{\mu_0-(1-\eta)eV}^{\mu_0+\eta eV}d\varepsilon\frac{\partial g}{\partial (eV)}(\varepsilon,T,eV)\Bigg\rvert_{\substack{V =0\\T=0}}
=g(\mu_0,0,0).\label{zeroordint1}
\end{align}
This yields $G_0=\frac{e^2}{h}\frac{\pi^2}{4}g(\mu_0,0,0)$. 
We use the zero bias forms of the coefficients $a_j$ and $b_j$, cf. Eqs. (\ref{genb1xsu2su2}) and (\ref{genb2xsu2su2}),   as well as the unitarity condition \eqref{TDOS cond1} found in the previous section.  Then it follows for an SU(2)$\otimes$SU(2) symmetric dot with 
lead coupling obeying  $\gamma_{L1}=\gamma_{L2}=\gamma_{L{\rm d}}$ and $\gamma_{L3}=\gamma_{L4}=\gamma_{L{\rm u}}$, 
\begin{align}
G_0&=\frac{2e^2}{h}\frac{\pi^2}{4}\frac{\Gamma}{\Gamma_d}
	\frac{	\gamma_{L{\rm d}}
		\gamma_{R{\rm d}}	}{\big[\frac{(\varepsilon_d-\mu_0)\pi}{2\Gamma_d}+\frac{1}{2}\big(\mathrm{ln}\big(\frac{W}{|{\cal E}|}\big)
		+	\frac{\Gamma_u}{\Gamma_d}\mathrm{ln}\big(\frac{W}{| {\cal E}+i{\Delta}|}\big)\big)\big]^2
		+ \big[\frac{\pi}{4}\big(1+\frac{\Gamma_u}{\Gamma_d}\big)-\frac{1}{2}\big(\varphi+\frac{\Gamma_u}{\Gamma_d}\varphi_{+
		}\big) \big]^2} \nonumber\\
&+\frac{2e^2}{h}\frac{\pi^2}{4}\frac{\Gamma}{\Gamma_u}
\frac{	\gamma_{L{\rm u}}
	\gamma_{R{\rm u}}
}{\big[\frac{(\varepsilon_u-\mu_0)\pi}{2\Gamma_u}+\frac{1}{2}\big(\mathrm{ln}\big(\frac{W}{|{\cal E}|}\big)
+	\frac{\Gamma_d}{\Gamma_u}\mathrm{ln}\big(\frac{W}{| {\cal E}-i{\Delta}|}\big)\big)\big]^2
	+ \big[\frac{\pi}{4}\big(1+\frac{\Gamma_d}{\Gamma_u}\big)-\frac{1}{2}\big(\varphi+\frac{\Gamma_d}{\Gamma_u}\varphi_{-
}\big) \big]^2
}
\end{align}
regardless of bias asymmetry. 
Using the unitary condition \eqref{TDOS cond1} this simplifies to 
\begin{align}
G_0&=\frac{2e^2}{h}
4(	\gamma_{L{\rm u}}\gamma_{R{\rm u}} - \gamma_{L{\rm d}}
		\gamma_{R{\rm d} })\frac{\pi}{2}\Gamma_u\nu_u (\mu_0)
+ \frac{2e^2}{h} 4\gamma_{L{\rm d}}
\gamma_{R{\rm d}}.
\end{align}
Notice that for the case of symmetric couplings, $\gamma_{\alpha i}=1/2$ the  value $G_0=2e^2/h$ is recovered. 
 Further, for large values of $\Delta$ the upper channel TDOS vanishes and hence $G_0\to \frac{2e^2}{h} 4\gamma_{L{\rm d}} \gamma_{R{\rm d}}$.

\subsection{The first order coefficient $G_1$   }
We proceed with the evaluation of the coefficient $G_1$ in the expansion of the differential conductance $G_{\rm diff}$. 
From \eqref{diffgtaylorsu2su2s} we need 
\begin{align}
&\frac{d^2}{d(eV)^2}\int_{\mu_0-(1-\eta)eV}^{\mu_0+\eta eV}d\varepsilon g(\varepsilon,T,eV)\Bigg\rvert_{\substack{V =0\\T=0}}=\Bigg\lbrace\underbrace{\left[\eta\frac{d}{d(eV)}g(\mu_L,T,eV) + (1-\eta)\frac{d}{d(eV)}g(\mu_R,T,eV) \right]}_{A} \nonumber\\
&+\underbrace{\bigg[\eta\frac{\partial }{\partial (eV)}g(\varepsilon,T,eV)\Bigg\rvert_{\varepsilon=\mu_L} + (1-\eta)\frac{\partial }{\partial (eV)}g(\varepsilon,T, eV)\Bigg\rvert_{\varepsilon=\mu_R} \bigg]}_{B}  +\underbrace{\int_{\mu_R}^{\mu_L}d\varepsilon\frac{\partial^2g}{\partial (eV)^2}(\varepsilon,T,eV)}_{C} 
 \Bigg\rbrace\Bigg\rvert_{\substack{V=0\\T=0}}. \label{no2expanda}
\end{align}
The term $C$ vanishes in the limit $V=0$. For the remaining terms we introduce the voltage asymmetries $\eta_L=\eta$ and $\eta_R=-1+\eta$. Further, we assign to the lead $\alpha$ the values $\alpha=L/R=\pm 1$. Then  we get the compact forms 
 \begin{align}
 A=\sum_\alpha \alpha \eta_\alpha \frac{d}{d(eV)} g(\mu_\alpha,T,eV)\Bigg\rvert_{\substack{V=0\\T=0}}, \qquad B=\sum_\alpha \alpha \eta_\alpha \frac{\partial}{\partial(eV)} g(\varepsilon,T,eV)\Bigg\rvert_{\substack{\varepsilon=\mu_\alpha\\V=T=0}}.
  \label{a+b}
 \end{align}
We start by evaluating the term $A$, which involves total derivatives with respect to the bias voltage of the kind 
\begin{align}
\frac{d}{d(eV)}\frac{1}{a_j^2(\mu_\alpha,T,eV)+b_j^2(\mu_\alpha,T,eV)}&=-2\frac{a_j\frac{d}{d(eV)}a_j(\mu_\alpha,T,eV)+b_j\frac{d}{d(eV)}b_j(\mu_\alpha,T,eV)}{\bigg( a_j^2(\mu_\alpha,T,eV)+b_j^2(\mu_\alpha,T,eV)\bigg)^2} , \label{dgdev}
\end{align}
Using relation \eqref{deldelepsi0} in the form  
\begin{align}
\frac{d}{d(eV)}\Psi\left(\frac{1}{2}+\frac{{\cal E}- i{\Delta_{ji}}}{2\pi k_B T}+i\frac{\mu_\alpha -\mu_\beta }{2\pi k_BT} \right)\Bigg\rvert_{\substack{V=0\\T=0}}
=i\frac{\eta_\alpha-\eta_\beta}{{\cal E} -i\Delta_{ji} }, 
\end{align}
we find
\begin{align}
\frac{d}{d(eV)}a_j(\mu_\alpha,T,eV)\Bigg\rvert_{\substack{V=T=0}}&= 
-\frac{\pi}{2\Gamma_j}\eta_\alpha -\frac{1}{2}\sum_{ i=T_j, C_j}\frac{\Gamma_i}{\Gamma_j}\sum_\beta\gamma_{{\beta} i}
\frac{\eta_\alpha -\eta_\beta}{\vert {\cal E}-i{\Delta_{ji}}\vert} \sin \varphi_{ji},\nonumber\\
\frac{d}{d(eV)}b_j(\mu_\alpha,T,eV)\Bigg\rvert_{\substack{V=T=0}}&= 
 -\frac{1}{2}\sum_{ i=T_j, C_j}\frac{\Gamma_i}{\Gamma_j}\sum_\beta\gamma_{{\beta} i}
\frac{\eta_\alpha -\eta_\beta}{\vert {\cal E}-i{\Delta_{ji}}\vert} \cos \varphi_{ji}.
\label{dadev}
\end{align}
Similarly, for the partial derivatives involved in the $B$ term we obtain
\begin{align}
\frac{\partial}{\partial(eV)}a_j(\varepsilon,T,eV)\Bigg\rvert_{\substack{\varepsilon=\mu_\alpha\\V=T=0}}&= 
 -\frac{1}{2}\sum_{ i=T_j, C_j}\frac{\Gamma_i}{\Gamma_j}\sum_\beta\gamma_{{\beta} i}
\frac{ (-\eta_\beta)}{\vert {\cal E}-i{\Delta_{ji}}\vert} \sin \varphi_{ji},\nonumber\\
\frac{\partial}{\partial(eV)}b_j(\mu_\alpha,T,eV)\Bigg\rvert_{\substack{\varepsilon=\mu_\alpha\\V=T=0}}&= 
-\frac{1}{2}\sum_{ i=T_j, C_j}\frac{\Gamma_i}{\Gamma_j}\sum_\beta\gamma_{{\beta} i}
\frac{(-\eta_\beta)}{\vert {\cal E}-i{\Delta_{ji}}\vert} \cos \varphi_{ji}.
\label{partialdadev}
\end{align}
Notice that the above derivatives were calculated at zero bias and do not involve the shift $\kappa$. 
On using the relation
\begin{align}
D_i&\equiv\sum_\alpha \alpha\eta_\alpha \sum_\beta \gamma_{\beta i}(\eta_\alpha-2\eta_\beta)=-\sum_\beta \beta \gamma_{\beta i} =\gamma_{{\rm R}i}-\gamma_{{\rm L}i} , \label{Diff}
\end{align}
we then obtain from Eqs. \eqref{a+b} together with \eqref{dadev} and \eqref{partialdadev}, 
\begin{align}
A+B&= \sum_j\frac{\Gamma}{\Gamma_j}\gamma_{{\rm L}j}\gamma_{{\rm R}j}
\frac{(-2)}{\bigg( a_j^2(\mu_0,0,0)+b_j^2(\mu_0,0,0)\bigg)^2} \nonumber\\
\times &
\Big\{a_j(\mu_0,0,0)\Big[-\frac{\pi}{2\Gamma_j}\sum_\alpha\eta_\alpha -\frac{1}{2}\sum_{ i=T_j, C_j}\frac{\Gamma_i}{\Gamma_j}
\frac{D_i}{\vert {\cal E}-i{\Delta_{ji}}\vert} \sin \varphi_{ji}
\Big]+b_j(\mu_0,0,0)\Big[ -\frac{1}{2}\sum_{ i=T_j, C_j}\frac{\Gamma_i}{\Gamma_j}
\frac{D_i}{\vert {\cal E}-i{\Delta_{ji}}\vert} \cos \varphi_{ji}
\Big]\Big\}, \label{a+bcompact}
\end{align}
from which $G_1=\frac{e^2}{\hbar}\frac{\pi}{8}(A+B)$ immediately follows. 
The contribution proportional to 
$a_j(\pi/2\Gamma_j)\sum_\alpha\eta_\alpha$ is usually neglected in the Kondo regime. As a result, $G_1$ becomes  independent of the bias asymmetries $\eta_\alpha$. Further, since $D_i=0$ in a symmetric set up with $\gamma_{\beta i}=1/2$, a non vanishing linear term $G_1$ requires that at least one $\gamma_{{\rm L} i}\neq \gamma_{{\rm R}i}$. 
With the aim of revealing some key behaviours for $\Delta \gg k_{\rm B}T_K$ in the crossover regime, we use that  $a_d\ll a_u$ for large $\Delta$. Then, we can approximate  
	\begin{align}
	G_1&=\frac{e^2}{h}\frac{\pi}{4}
	\frac{\frac{\Gamma}{\Gamma_d}\gamma_{{\rm L}d}\gamma_{{\rm R}d}}{\big[ a_d^2(\mu_0,0,0)+b_d^2(\mu_0,0,0)\big]^2}
	\bigg[(\gamma_{Rd}-\gamma_{Ld})\bigg(a_d(\mu_0,0,0)\frac{\sin \varphi}{\vert {\cal E}\vert}+b_d(\mu_0,0,0)\frac{\cos \varphi}{\vert {\cal E}\vert}\bigg)\nonumber\\&+(\gamma_{Lu}-\gamma_{Ru})\bigg(a_d(\mu_0,0,0)\frac{\frac{\Gamma_{u}}{\Gamma_d}\cos \varphi_{+}}{\vert {\cal E}+i\Delta\vert}+b_d(\mu_0,0,0)\frac{\frac{\Gamma_{u}}{\Gamma_d}\cos \varphi_{+}}{\vert {\cal E}+i\Delta\vert}\bigg)\bigg]+\mathcal{O}\left((k_{\rm{B}}T_K/\Delta)^4\right).
	\end{align}
Finally we turn to the SU(4) case, characterized by $\Delta=0$, $\Gamma_j=\Gamma$, 
 $a_j(\mu_0,0,0)=a$, $b_j(\mu_0,0,0)=b$, and $\gamma_{\alpha i}=\gamma_\alpha$. Then 
Eq. \eqref{a+bcompact} yields (upon neglecting the term proportional to $a/\Gamma$)
\begin{align}
G_1=&\frac{2e^2}{h}
\frac{\pi^2\gamma_{L}\gamma_{R}\big( \gamma_{R}-\gamma_{L}\big)}{\Big(a^2(\mu_0,0,0))+b^2(\mu_0,0,0)\Big)^2}
\Bigg[-\mathrm{ln}\Big(\frac{2\vert{\cal E}\vert}{k_BT_K}\Big)
\left(\frac{\mathrm{sin}\varphi}{\vert{\cal E}\vert}\right)+ \left(\frac{\pi}{2}-\varphi\right) \left(\frac{\mathrm{cos}\varphi}{\vert {\cal E}\vert}\right)\Bigg]. \label{linordsu4}
\end{align}
Comparing the term in the bracket with the expression  ~\eqref{su4cond2k}, accounting  for the shift $\kappa$ of the location of the TDOS maximum from the  energy $\mu_0$, we see that the two coincide if   the shift $\kappa$ is neglected.  Thus, as a consequence of \eqref{su4cond2k}, the term $G_1$ would vanish if the SU(4) shift $\kappa$ is neglected. 
 Lastly, the SU(2) case has no linear term irrespective of the asymmetries. Hence the zero-bias peak does not shift at all. The conclusions of our theory match that of Ref. \citen{PhysRevB.80.155322}. \\

\section{Kondo transitions: TDOS and self-energy signatures}\label{Kondotrans}
In this section we shortly review the mechanism leading to Kondo resonances. 
\subsection{Self-energy resonances}
\begin{figure}[h!]
		\includegraphics[width=\linewidth]{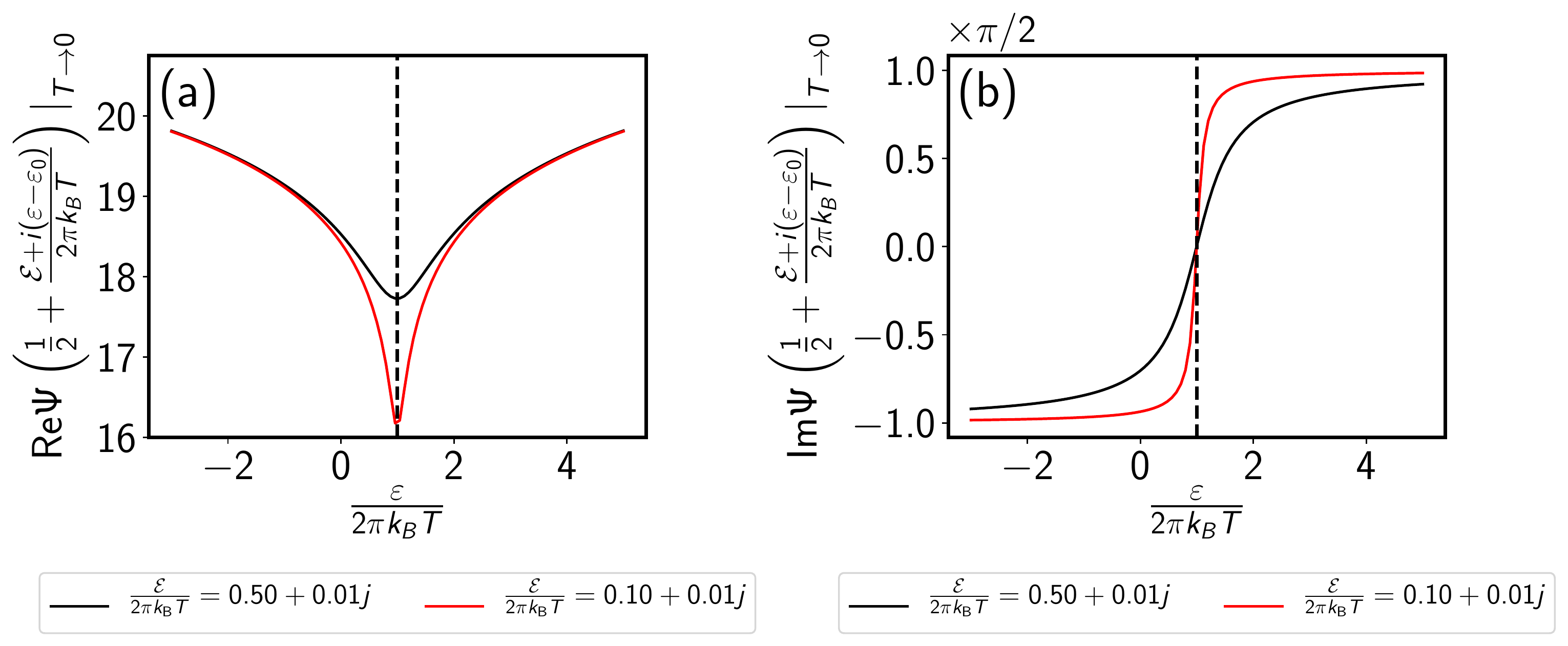}
	\caption{Mechanism of resonances in the TDOS. Resonances in the Kondo regime are related to the low temperature behavior of the constituent digamma functions in the self-energies. These occur at $\varepsilon=\varepsilon_0 -{\rm Im}{\cal E}$, where the real part has a dip while the imaginary part has a phase change of $\pi$. 
		The resonance features are sharp  for small ${\cal E}/2\pi k_{\rm B}T$. For the simulation we choose $\varepsilon_0/2\pi k_{\rm B}T=1$, corresponding to the location of the vertical dashed line in panels (a) and (b).   }
	\label{digammabehav}
\end{figure}

Since for the model considered in this work each channel $j$ contributes independently to the current, it is sufficient to analyze the current per channel
\begin{equation}
I_j = \frac{e}{\hbar}\int_{-\infty}^{\infty}\frac{\Gamma_{L,j} \Gamma_{R,j}}{\Gamma_{L,j}+\Gamma_{R,j}}\nu_j(\varepsilon)(f_L(\varepsilon)-f_R(\varepsilon))d\varepsilon \;. \label{currentformula}
\end{equation}
Resonant features of the differential conductance $G_\mathrm{diff}=\sum_j\frac{dI_j}{dV}$ are thus related to resonant features of the associated channel TDOS.  From~\eqref{TDOS}, the denominator of the channel TDOS $\nu_j(\varepsilon)$ bears the form,
\begin{align}
\left[(\varepsilon_j-\varepsilon)+\Gamma\mathrm{Re}\Sigma_j(\varepsilon)\right]^2+\left[\Gamma\mathrm{Im}\Sigma_j(\varepsilon)\right]^2. \label{tdosden}
\end{align}
Thus, resonances are associated to values of the parameter $\varepsilon$ which minimize the denominator \eqref{tdosden}. Further, these values have to lie in the energy window set by the difference of Fermi functions in \eqref{currentformula}. Hence, in the sequential tunneling regime where 
$\varepsilon_j$ lies in the transport window (e.g. $\mu_L>\varepsilon_j>\mu_R$ for $eV>0$), the 
TDOS has a peak at $\varepsilon=\varepsilon_j$, i.e. at the charge transfer peak. In the off-resonant regime, $\mu_L, \mu_R \gg \varepsilon_j$,  the nature and number of resonances crucially depends on the energy dependence of the self-energy $\Sigma_j(\varepsilon)$, and in turn, according 
to \eqref{Selfeform}, on the energy dependence of  digamma functions of the form $\Psi\left(\frac{1}{2} + \frac{{\cal E}}{2\pi k_BT} + i\frac{(\varepsilon-\varepsilon_0)}{2\pi k_BT}\right)$ \cite{Smirnov2011a}. 
Such a behavior is shown in in Fig. \ref{digammabehav}. 
The real part of the digamma function has a dip at $\varepsilon =  \varepsilon_0 -{\rm Im }{\cal E}$, and correspondingly the imaginary part has a  change of $\pi$. We refer to these features henceforth as "resonant features". 
%
%
The sharpness of the peak and the abruptness of the sudden drop depend on the temperature $k_{\rm B}T$ as well as on the magnitude of ${\cal E}$. For $T\to 0$, from Eq. \eqref{psilargezlim}, a constituent digamma function in the self-energy Eq. \eqref{Selfeform} is found to have the form, 
\begin{align}
\Psi\left(\frac{1}{2} + \frac{{\cal E}}{2\pi k_BT} + i\frac{(\varepsilon-\varepsilon_0)}{2\pi k_BT}\right)\xrightarrow{T\to 0}\mathrm{ln}\left(\frac{\vert{\cal E}+ i(\varepsilon-\varepsilon_0)\vert}{2\pi k_BT}\right)+i\underbrace{\mathrm{atan}\left(\frac{\mathrm{Im}({\cal E}+ (\varepsilon-\varepsilon_0))}{\mathrm{Re}({\cal E})}\right)}_{\xrightarrow{\mathrm{Re}{\cal E}\to 0}\mathrm{sign}(\varepsilon-(\varepsilon_0+\mathrm{Im}{\cal E}))\frac{\pi}{2}}\label{digambeh}
\end{align}
which has a logarithmic dip in its real part at $\varepsilon=\varepsilon_0-\mathrm{Im}{\cal E}$, where the argument $\vert{\cal E}+i(\varepsilon-\varepsilon_0)\vert$ is the smallest, and a corresponding rise of $\pi$ in its imaginary part. Additionally, at the tip of the dip, the real part assumes the value $\mathrm{ln}\left(\frac{\mathrm{Re}{\cal E}}{2\pi k_BT}\right)$, while the width of the resonant feature is of the order of $\mathrm{Re}{\cal E}$. Consequently, from Eqns. \eqref{Selfeform} and \eqref{digambeh}, it is clear that the real part of the self-energy exhibits a peak while the imaginary part has a sudden drop, leading to a Kondo peak in the TDOS. Further, a stronger peak in the real part and correspondingly, a sharper and/or larger drop in the imaginary part of the self-energies (i.e. stronger resonant feature), arising from a larger value of $\mathrm{Re}{\cal E}$, leads to a stronger peak in the TDOS. Clearly, the strength of the Kondo resonances depend strongly on $\mathrm{Re}{\cal E}$ while the location is slightly renormalized by $\mathrm{Im}{\cal E}$.
%
\subsection{$P-$transitions}
\begin{figure}[h!]
\renewcommand{\thesubfigure}{\Roman{subfigure}}
\centering
\begin{subfigure}{.49\textwidth}
  \centering
  \includegraphics[width=\linewidth]{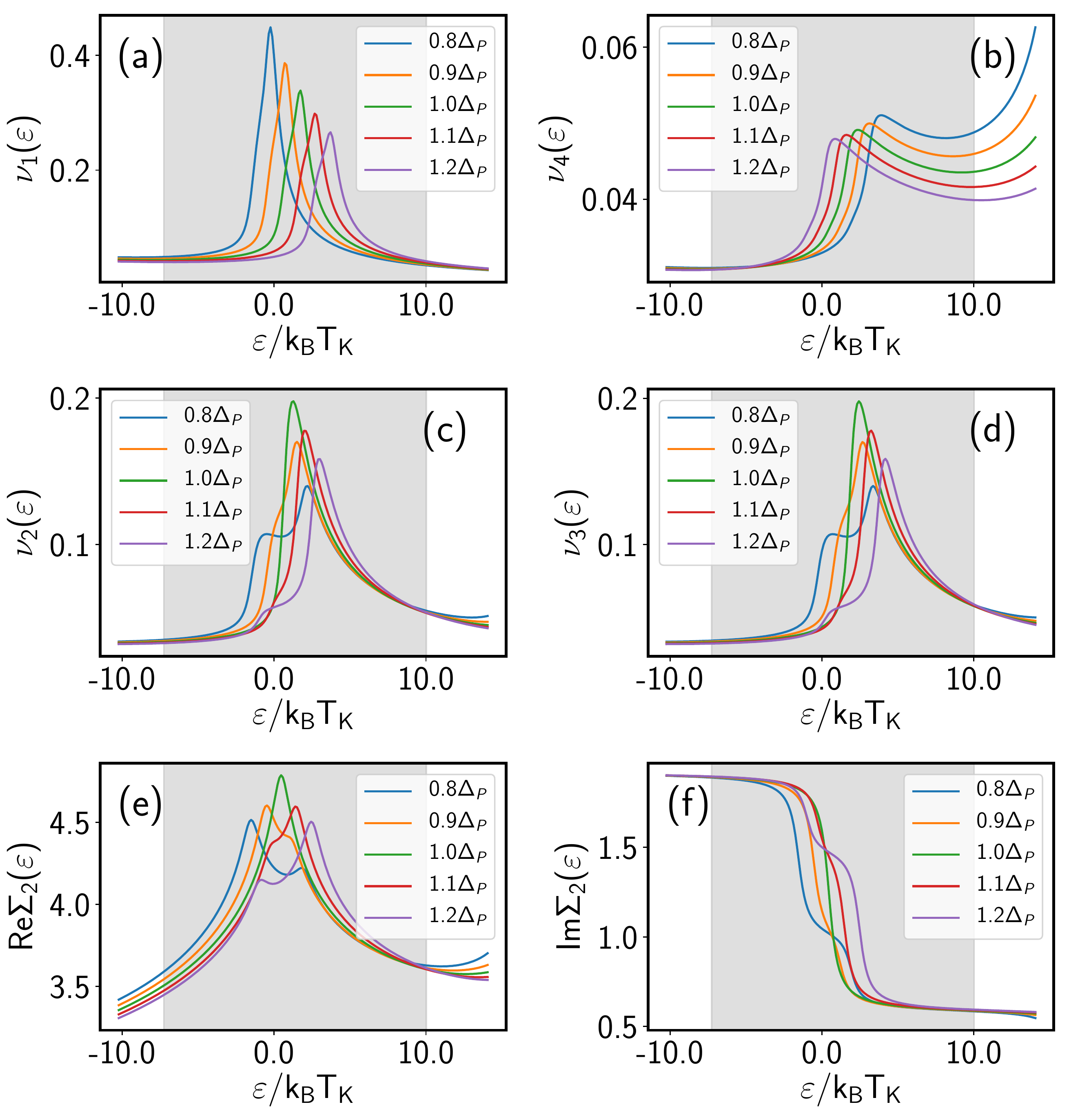}
\caption{\centering N=1}
\label{tcconcn1}
\end{subfigure}
\begin{subfigure}{.49\textwidth}
  \centering
\includegraphics[width=\linewidth]{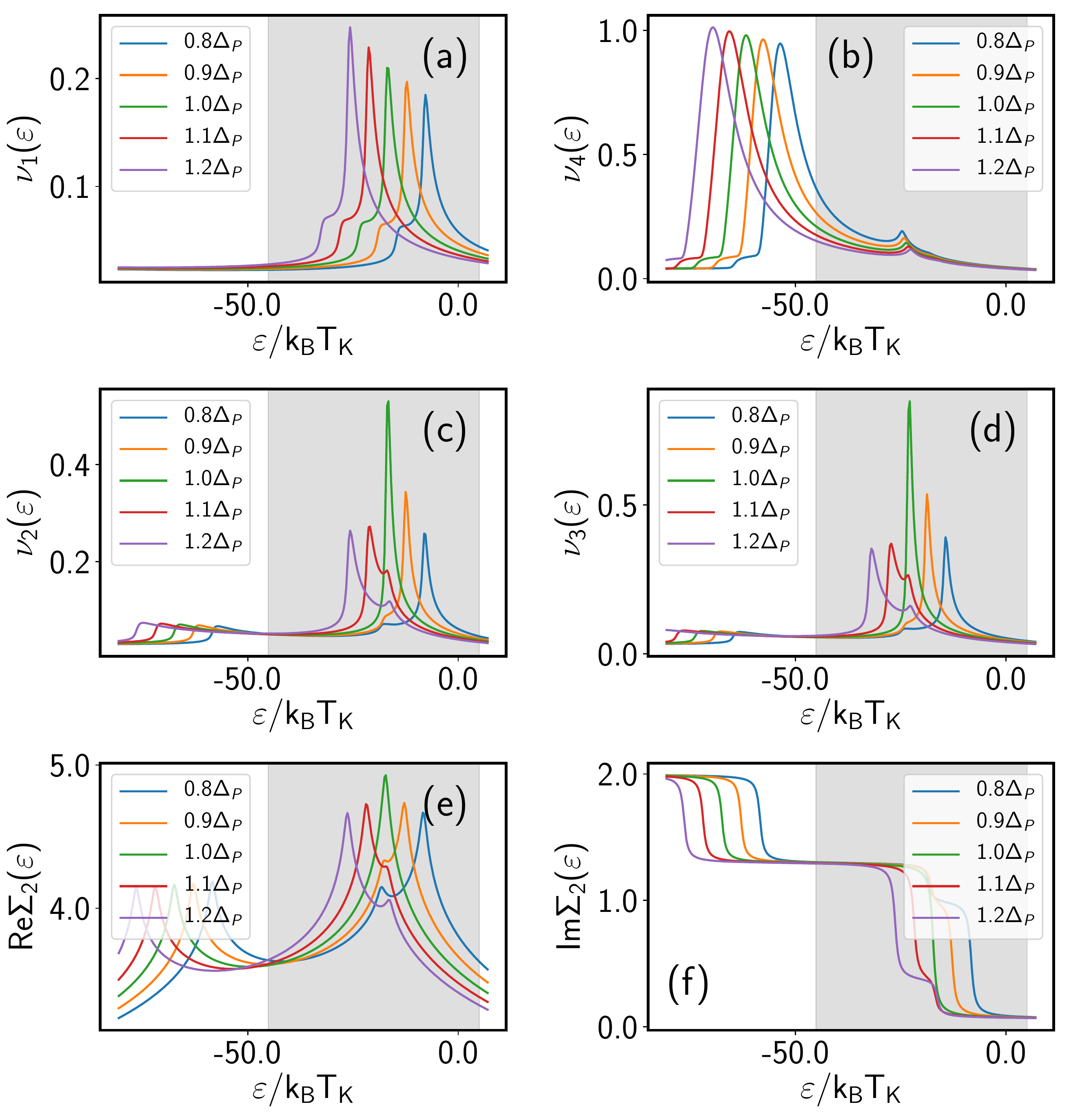}
\caption{\centering N=3}
\label{tcconcn3}
\end{subfigure}
\renewcommand{\thesubfigure}{\arabic{subfigure}}
\caption{TDOS signatures of the $P$-resonance. Channel density of states $\nu_j$, (a)-(d), and  self-energy $\Sigma_2$ (e), (f), evaluated at bias drops around the energy $\Delta_P$ of the $P$-resonance, for the case $N=1$, and  $N=3$ case. The gray stripe indicates the integration range set by the lead chemical potentials. At $eV=\Delta_P$ the channel density of states $\nu_2$ and $\nu_3$ are maximal. This is due  to a resonance of the associated self-energy, as illustrated in (e), (f) on the example of $\Sigma_2$. The magnetic field is $B=8.05$~T, and we set $\Gamma_u=\Gamma_d=\Gamma$.   }
\label{tcconc}
\end{figure}

 From  the above discussion and \eqref{Selfeform} it follows that the digamma functions, and thus the associated self-energies, display Kondo resonances anytime  $\varepsilon\approx \mu_\alpha+\Delta_{ji}$. This denotes a Kondo transition process between the levels $j\leftrightarrow i$ mediated by the lead $\alpha$.
  In particular, as discussed in the main text, the resonances of the self-energy $\Sigma_2$ at $\varepsilon=\mu_R + \Delta_{21}$ and $ \varepsilon=\mu_L + \Delta_{24}$ merge in a single resonance when the bias drop  $eV$ is such that $\mu_L-\mu_R=\Delta_{42} + \Delta_{21}= \Delta_{41}=\Delta_P$. Likewise for the self-energy $\Sigma_3$. This effect was shown in Fig. 4 of the main text for the case of occupation $N=1$. This behavior together with the one for occupation $N=3$ is shown in  \ref{tcconc} for the same parameter set.
In Fig.~\ref{tcconc}, the channel tunneling density of states $\nu_j$ and the self-energy $\Sigma_2$ of the level two are shown. Similar to the $N=1$ case, the tunneling density of states $\nu_2$ and $\nu_3$ show a peak when the applied bias drop matches the energy $\Delta_P$ associated with the $P$-transition. 
\subsection{Effect of tunneling coupling asymmetries}
\label{Kondoasymm}
From the above discussion it is clear that the condition for a $P$-resonance, $eV=\Delta_P$,  is independent of  the bias asymmetry $\eta$ as well of the coupling asymmetries. However, the magnitude of the resonance, and hence whether the resonance is visible or not, is ruled also by asymmetries. For this we start looking at the role coupling asymmetries play in the magnitude of the drop of   Im$\Sigma_2$ and Im$\Sigma_3$ at a $P$-transition.
For clarity, let us consider the behavior of the self-energy $\Sigma_2$. Further, we consider for simplicity the situation in which $\Gamma_u=\Gamma_d=\Gamma$.
  From \eqref{Selfeform}, we have that 
\begin{align}
&\Sigma_2(\varepsilon)=\frac{1}{\pi}\Bigg[2\mathrm{ln}\left(\frac{W}{2\pi k_{\rm B}T}\right)+2i\frac{\pi}{2}-\gamma_{L{\rm u}}\Psi\left(\frac{1}{2}+\frac{{\cal E}}{2\pi k_{\rm B}T}-\frac{i\mu_L}{2\pi k_{\rm B}T}+\frac{i\varepsilon}{2\pi k_{\rm B}T}-\frac{i\Delta_{2,C_2}}{2\pi k_{\rm B}T}\right)\nonumber \\&-\gamma_{R{\rm u}}\Psi\left(\frac{1}{2}+\frac{{\cal E}}{2\pi k_{\rm B}T}-\frac{i\mu_R}{2\pi k_{\rm B}T}+\frac{i\varepsilon}{2\pi k_{\rm B}T}-\frac{i\Delta_{2,C_2}}{2\pi k_{\rm B}T}\right)-\gamma_{L{\rm d}}\Psi\left( \frac{1}{2}+\frac{{\cal E}}{2\pi k_{\rm B}T}-\frac{i\mu_L}{2\pi k_{\rm B}T}+\frac{i\varepsilon}{2\pi k_{\rm B}T}-\frac{i\Delta_{2,T_2}}{2\pi k_{\rm B}T}\right)\\&-\gamma_{R{\rm d}}\Psi\left( \frac{1}{2}+\frac{{\cal E}}{2\pi k_{\rm B}T}-\frac{i\mu_R}{2\pi k_{\rm B}T}+\frac{i\varepsilon}{2\pi k_{\rm B}T}-\frac{i\Delta_{2,T_2}}{2\pi k_{\rm B}T}\right)  \Bigg].\nonumber
\end{align}
From the behavior of the digamma function as shown in Fig.~\ref{digammabehav}, it is clear that for very large negative values of $\varepsilon$ is Im$\Sigma_2(\varepsilon)=2$, see Fig.~\ref{tcconc}. In fact, at zero temperature we have,
\begin{align}
\lim_{\varepsilon\to-\infty}\mathrm{Im}\Sigma_2(\varepsilon)&=\frac{1}{\pi}\sum_{\alpha=L,R}\Big[\frac{\pi}{2}+\gamma_{\alpha {\rm u}}\frac{\pi}{2}+\gamma_{\alpha {\rm d}}\frac{\pi}{2} \Big]=\frac{1}{\pi}\Big[2\frac{\pi}{2}+\frac{\pi}{2}+\frac{\pi}{2}\Big]=2. 
\end{align}
At finite energies,  there are four parameter configurations for which the self-energy shows resonant features. Namely,
drops in $\mathrm{Im}\Sigma_{2}$ are found at: $\varepsilon=\mu_R+\Delta_{2,C_2}$, $\mu_R+\Delta_{2,T_2}$, $\mu_L+\Delta_{2,C_2}$ and $\mu_L+\Delta_{2,T_2}$. Note that, for $\mu_L-\mu_R=\Delta_P$, regardless of the asymmetries, $\mu_L+\Delta_{2,C_2}=\mu_R+\Delta_{2,T_2}$, which is fundamental to the $P-$transitions.
  The first resonant feature is located at $\varepsilon=\mu_R+\Delta_{2,C_2}$ for $eV>0$.  We get 
\begin{align}
\mathrm{Im}\Sigma_2(\varepsilon, 1st\text{ resonant feature crossed})&=2-\gamma_{R{\rm u}}=1+\gamma_{L{\rm u}}. \label{imsig21}
\end{align}
Similarly, after crossing three resonant features, barring the one located at the highest  energy,
\begin{align}
\mathrm{Im}\Sigma_2(\varepsilon, 3rd
\text{ resonant features crossed})&=1-\frac{1}{2}\left(\gamma_{R{\rm u}}+\gamma_{R{\rm d}}+\gamma_{L{\rm u}} -\gamma_{L {\rm d}}\right)=\gamma_{L{\rm d}}.  \label{imsig23}
\end{align}
Therefore, when the resonant features corresponding to the $T$- and $C$-transitions merge, the height of the drop in Im$\Sigma_2(\varepsilon)$ is given by 
\begin{align}
\Delta\mathrm{Im}\Sigma_2(\varepsilon, eV=\Delta_P)&=1+\gamma_{L{\rm u}}-\gamma_{L{\rm d}}.
\end{align}
 Clearly, this is directly dependent on the coupling asymmetry factors. Eqns.~\eqref{imsig21} and~\eqref{imsig23} may be verified by comparing with Fig.~\ref{tcconc}(f). Further, when the upper Kramers pair is more strongly coupled to the left lead than the lower Kramers pair, i.e., $\gamma_{L{\rm{u}}}>\gamma_{L{\rm d}}$, then $\Delta\mathrm{Im}\Sigma_2(\varepsilon, eV=\Delta_P)$ increases implying a stronger resonant feature. Consequently, from the discussion in Sec.~\ref{Kondotrans}, it is concluded that this strengthens the peak in the TDOS creating/strengthening the $P-$transition. 
Even though we can not give further analytical arguments, this suggests that a threshold value $\zeta_1$ exists for the difference between $\gamma_{L{\rm u}}$ and $\gamma_{L{\rm d}}$ above which a $P$-resonance is seen. 
Now, one may repeat the derivation performed above using the self-energy expressions for both the $N=1$ and $N=3$ cases, along with $eV=\pm \Delta_P$. This yields the table below, describing the parameter domains for obtaining a $P$-transition. The small positive constants $\zeta_{1/3}$ represent threshold values for the magnitude of the coupling asymmetry required to yield a $P$-transition.
\begin{table}[h!]
	\label{su2table}
	\begin{center}
		\begin{tabular}{||c c c||} 
			\hline 
			$N$&$\mu_L-\mu_R>0$&\qquad$\mu_L-\mu_R<0$\\
			\hline\hline
			1 &  $\gamma_{L{\rm u}}-\gamma_{L{\rm d}}>\zeta_1$ & \qquad$\gamma_{L{\rm d}}-\gamma_{L{\rm u}}>\zeta_1$\\
			3 & $\gamma_{L{\rm d}}-\gamma_{L{\rm u}}>\zeta_3$ & \qquad$\gamma_{L{\rm u}}-\gamma_{L{\rm d}}>\zeta_3$\\
			\hline
		\end{tabular}
	\end{center}
	\caption{The domains in the coupling asymmetry parameter space yielding a $P-$transition. Notice that  $\zeta_{1/3}>0$.}
\end{table}

\begin{figure}[h!]
	\includegraphics[width=\linewidth]{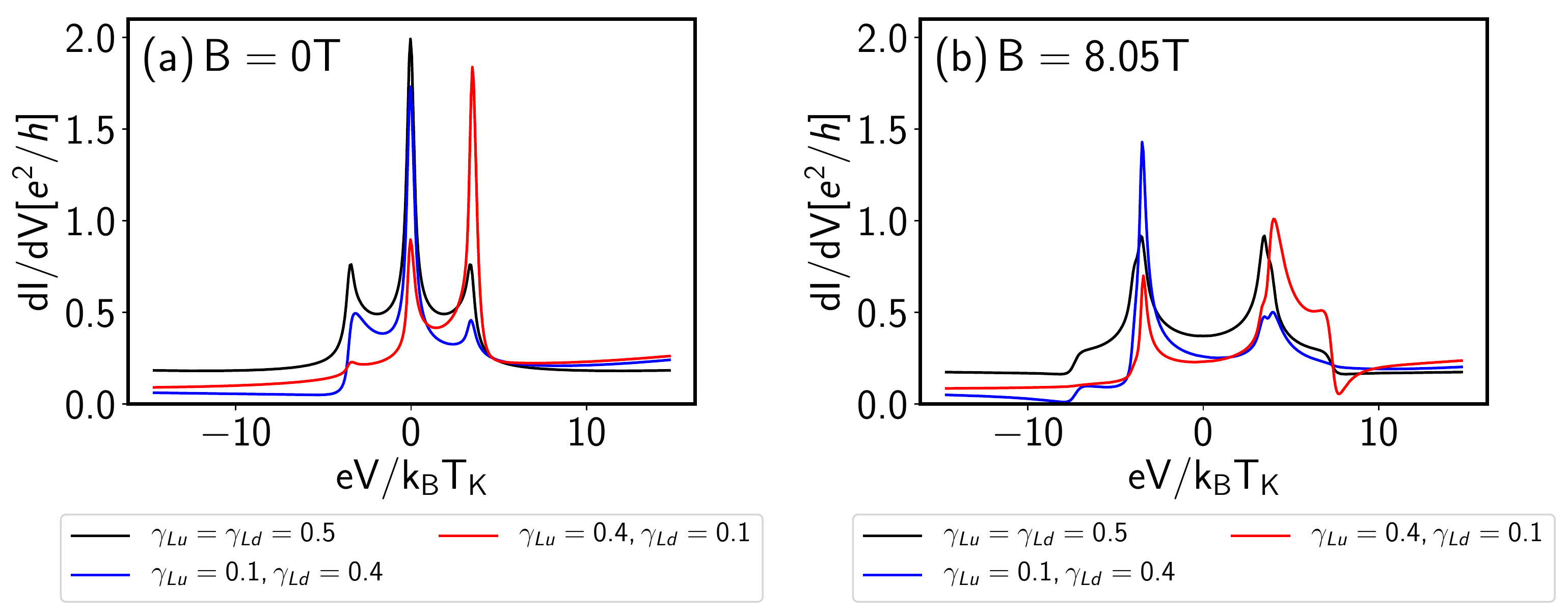}
\caption{Differential conductance for the $N=1$ case for (a) $B=0$~T and (b) $B=8.05$~T for various coupling asymmetries.  The remaining parameters are the one used to match the $N=1$ experimental data.  The red curve in (b) with $\gamma_{L{\rm u}}>\gamma_{L{\rm d}}$ shows a $P$-transition at $\mu_L-\mu_R>0$. }
\label{n1asymm}
\end{figure}

The impact of the tunneling coupling asymmetry  on the $P$-transition is shown in Fig.~\ref{n1asymm} for the $N=1$ case. Fig.~\ref{n1asymm}(a) is taken as a reference as it shows the case of zero magnetic field which does not show a $P$-transition.  
For  Fig.~\ref{n1asymm}(b) a large value of the magnetic field, $B=8.05$~T, is chosen. 
It is clear that  equal tunnel couplings to both the Kramers pairs do not produce a $P-$transition. For the case  $\gamma_{L {\rm u}}>\gamma_{L {\rm d}}$ and negative bias voltages  the KEA theory yields a small peak  at the position of the $P$-transition - with an adjacent dip - in the differential conductance. The dip is not seen in the experiment. We attribute this discrepancy to the lack of some cotunneling contributions in the KEA theory which can become relevant at high energies or to the strong impact of the Coulomb blockade peak which in the experiment might screen the dip. 
\subsection{Bias drop asymmetries}
\label{Kondoasymmbias}
\begin{figure}[h!]
  \centering
  \includegraphics[width=0.9\linewidth]{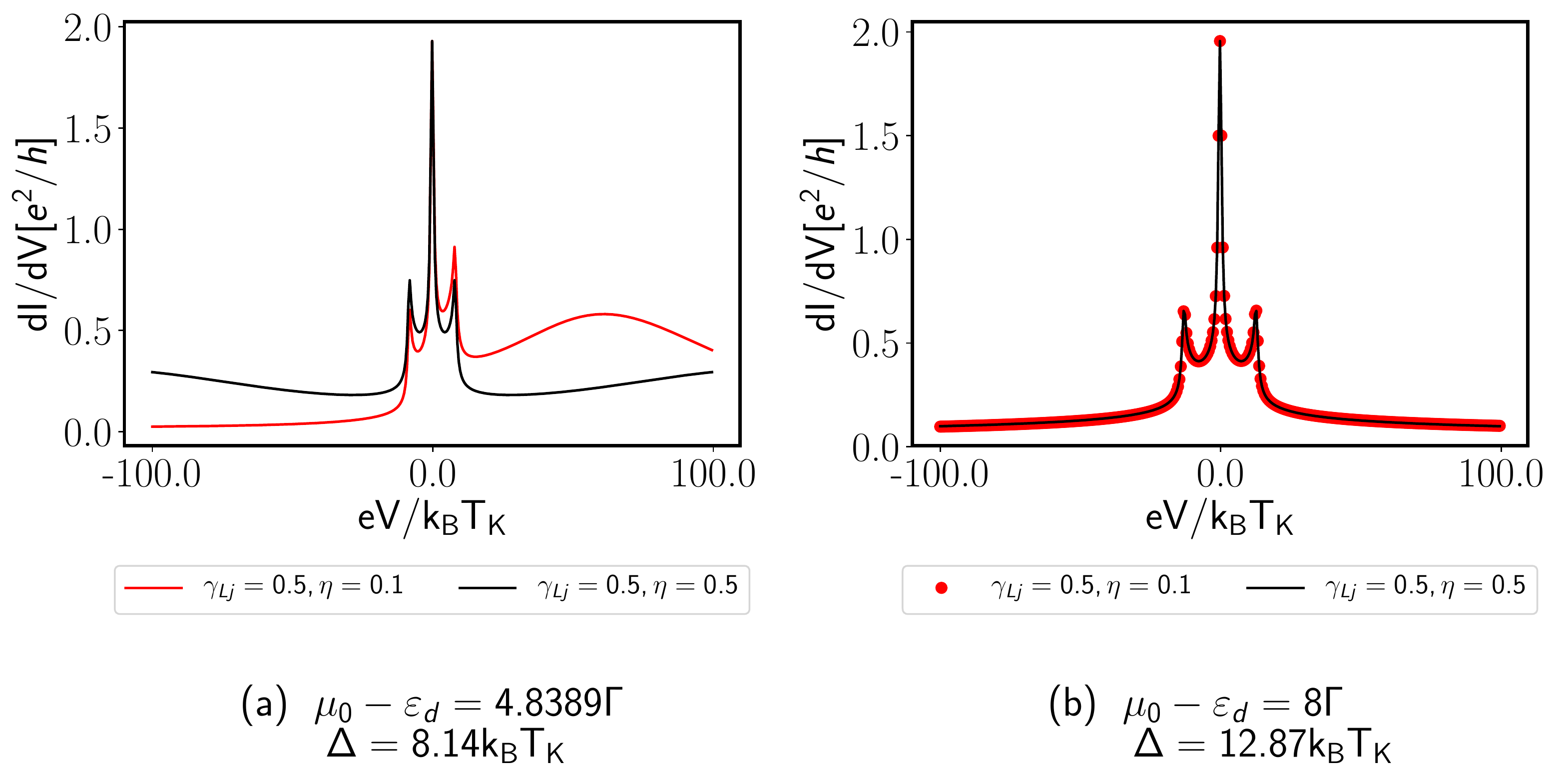}
\caption{Interplay of bias asymmetries and charge transfer peak on the differential conductance.  The  $N=1$ case at $B=0$~T is shown for two distinct values of the asymmetry parameter $\eta$ for (a)   $\mu_0-\varepsilon_d=4.8389\Gamma$, and (b)  $\mu_0-\varepsilon_d=8\Gamma$. The remaining dot parameters are the ones used to match the $N=1$ experimental data. Notice  the negligible effect of bias asymmetries  on the Kondo resonances for the parameter set in (b).   }
\label{n1asymmbias}
\end{figure}

Bias drop asymmetries have a much smaller and much less dramatic effect on the Kondo resonances compared to the coupling asymmetries as they primarily affect the high-energy/high-bias charge-transfer resonances. However, for a small value of $\mu_0-\varepsilon_d$, the Lorentzian tails of the charge-transfer peaks may extend all the way to the low-bias/near zero-bias regime and skew the Kondo peak amplitudes, as seen in Fig.~\ref{n1asymmbias}(a). 
For larger differences the Kondo resonances remain untouched, as seen  in Fig.~\ref{n1asymmbias}(b) where  $\mu_0-\varepsilon_d=8\Gamma$ is chosen. Here, even in the presence of coupling asymmetries the bias asymmetries remain ineffective.
This result can be understood by inspection of the self energy $\Sigma_j$ and the associated channel TDOS $\nu_j$. From ~\eqref{Selfeform} the former may be written in the presence of bias asymmetries as  (here for simplicity $\Gamma=\Gamma_i=\Gamma_u=\Gamma_d$ was chosen)
\begin{align}
\Sigma_j(\varepsilon)&=\frac{1}{\pi}\sum_{i=T_j,C_j}\Bigg[\mathrm{ln}\left(\frac{W}{2\pi k_BT}\right)+i\frac{\pi}{2}-\gamma_{Li}\Psi\Bigg(\frac{1}{2}+\frac{{\cal E}}{2\pi k_BT}-i\frac{\overbrace{eV/2+\left(\eta-1/2\right)eV}^{\mu_L}-\varepsilon+\Delta_{ji}}{2\pi k_BT} \Bigg)\nonumber\\
&-\gamma_{Ri}\Psi\Bigg(\frac{1}{2}+\frac{{\cal E}}{2\pi k_BT}-i\frac{\overbrace{-eV/2+\left(\eta-1/2\right)eV}^{\mu_R}-\varepsilon+\Delta_{ji}}{2\pi k_BT} \Bigg) \Bigg]\nonumber\\
&=\frac{1}{\pi}\sum_{i=T_j,C_j}\Bigg[\mathrm{ln}\left(\frac{W}{2\pi k_BT}\right)+i\frac{\pi}{2}-\gamma_{Li}\Psi\Bigg(\frac{1}{2}+\frac{{\cal E}}{2\pi k_BT}-i\frac{\frac{eV}{2}-\varepsilon'+\Delta_{ji}}{2\pi k_BT} \Bigg)\nonumber\\
&-\gamma_{Ri}\Psi\Bigg(\frac{1}{2}+\frac{{\cal E}}{2\pi k_BT}-i\frac{-\frac{eV}{2}-\varepsilon'+\Delta_{ji}}{2\pi k_BT} \Bigg) \Bigg]
:=\Sigma^{\rm sym}_j(\varepsilon'). \label{biasasymmselfeshift}
\end{align}
Here we have made the substitution $\varepsilon'=\varepsilon-\left(\eta-\frac{1}{2}\right)eV$, which recasts the self-energy in a bias-symmetric form without altering its features. In contrast, for the  TDOS one finds from~\eqref{TDOS} and~\eqref{biasasymmselfeshift}, 
\begin{align}
\nu_j(\varepsilon)&=\frac{1}{2\pi\Gamma_j}\frac{1}{\left(\frac{\varepsilon_j-\varepsilon}{\Gamma_j }+\mathrm{Re}\Sigma_j(\varepsilon)\right)^2+\left( \mathrm{Im}\Sigma_j(\varepsilon)\right)^2}\nonumber\\
&=\frac{1}{2\pi\Gamma_j}\frac{1}{\left(\frac{\varepsilon_j-\varepsilon'}{\Gamma_j }-\frac{(\eta-1/2)eV}{\Gamma_j }++\mathrm{Re}\Sigma_j^{\rm sym}(\varepsilon')\right)^2+\left( \mathrm{Im}\Sigma_j^{\rm sym}(\varepsilon')\right)^2}.
\end{align}
This shows that when the $\eta$-dependent contribution in the bracket can be neglected, as for the parameter set in Fig. ~\ref{n1asymmbias}(b), the replacement $\varepsilon \to \varepsilon'$ just amounts to a rigid shift of   the integration  window which does not affect the differential conductance.
\section{Matching between theory and experiment}
From the experimental data it is possible to extract $T_K$, $\Gamma$ and $\Delta$. Specifically:
i) The Kondo temperature $T_K(\Delta)$ is estimated from the width of the zero-bias peaks according to the approximate relation \cite{Pletyukhov2012} 
$G_{\mathrm{diff}}(eV/k_{\rm B}T_K(\Delta))=(2/3)G_{\mathrm{diff}}(eV=0)$. We find 
$T_{K1}=1$ K and $T_{K3}=0.37$ K for valley $N=1$ and $N=3$, respectively.
 ii) The Kramers splitting $\Delta\simeq0.7$ meV is extracted from the distance between the inelastic peaks of the differential conductance $G_{\mathrm{diff}}$. iii) The average linewidth $\Gamma\simeq 2.44$ meV is extracted from a Lorentzian fit to the contribution of the charge transfer peaks, as discussed in subsection \ref{gammaest} below. This in turn yields an estimate for the ratios $k_{\rm B}T_{K1}/\Gamma\simeq 0.025$, $k_{\rm B}T_{K3}/\Gamma\simeq 0.013$. 
 
 Due to our assumption of infinite charging energy, $U\to\infty$, the theoretical Kondo temperature is given by  
 \begin{align}
 k_{\rm B}T_K(\Delta)=f(\Delta)k_{\rm B}T_K=f(\Delta) \left[2W\mathrm{exp}\left(-\pi\frac{(\mu_0-\varepsilon_d)}{2\Gamma}\right)\right],
 \label{Kondosu22}
 \end{align}
 where the ratio $(\mu_0-\varepsilon_d)/\Gamma$ as well as the prefactor $f(\Delta)W$ cannot be directly extracted from the experiment.  At low bias  this is  not important, since the relevant transport quantities show a universal behavior which is not affected by bias and tunneling asymmetries, as seen in Fig.~\ref{n1asymmbias}. I.e.,   the experimentally measured and theoretically evaluated  differential conductance collapse onto a universal curve when they are scaled by the respective  Kondo temperatures. 
At larger bias, $\vert eV \vert \approx\Delta$, the charge transfer peak and the asymmetries may affect the differential conductance and thus a reasonable range of $(\mu_0-\varepsilon_d)/\Gamma$ has to be chosen for the simulations as well.   In our theory the bandwidth $W$ is a free parameter. By fixing it to be $W/\Gamma=100$, we find  $\mu_0-\varepsilon_d=5.3643\Gamma$ using the procedure described in  the following subsection \ref{dotlevelcalc}. 
\subsection{Estimate of the ratio $(\mu_0-\varepsilon_d)/\Gamma$  }
\label{dotlevelcalc}
Given the experimentally obtained values of the Kondo temperature $T_K$, of the splitting $\Delta$,  and of the tunnel coupling $\Gamma$, the calculation of $\varepsilon_d-\mu_0$ requires an iterative procedure. 
Since the prefactor $f(\Delta)$ in Eq. \eqref{Kondosu22} is unknown, we first fix $W/\Gamma$ and then proceed iteratively:
\begin{itemize}
\item An initial  value $f=f_1$ is assumed. By imposing $T_K({\rm theory})=T_K({\rm experiment})$ a value for $(\mu_0-\varepsilon_d)/\Gamma$ is found.
\item Using these values and $\Delta$,  we find ${\cal E}$ from the unitary conditions Eqns.~\eqref{TDOS cond1} and~\eqref{su4cond2k}. 
\item We obtain the Kondo temperature from the linear conductance using $G_0(T_K(\Delta))=G_0(T=0)/2 $.  This gives us $f=f_2 $. 
\item Steps 2 and 3 are repeated until convergence for  $f$.
\item This value of $f$ is used again in step 1 to get a new value for $\mu_0-\varepsilon_d$.
\item Steps 1-5 are repeated until convergence. 
\end{itemize}

\subsection{Contribution of the charge-transfer peaks to the Kondo resonances}
\label{gammaest}
\begin{figure}[h!]
  \includegraphics[width=0.9\linewidth]{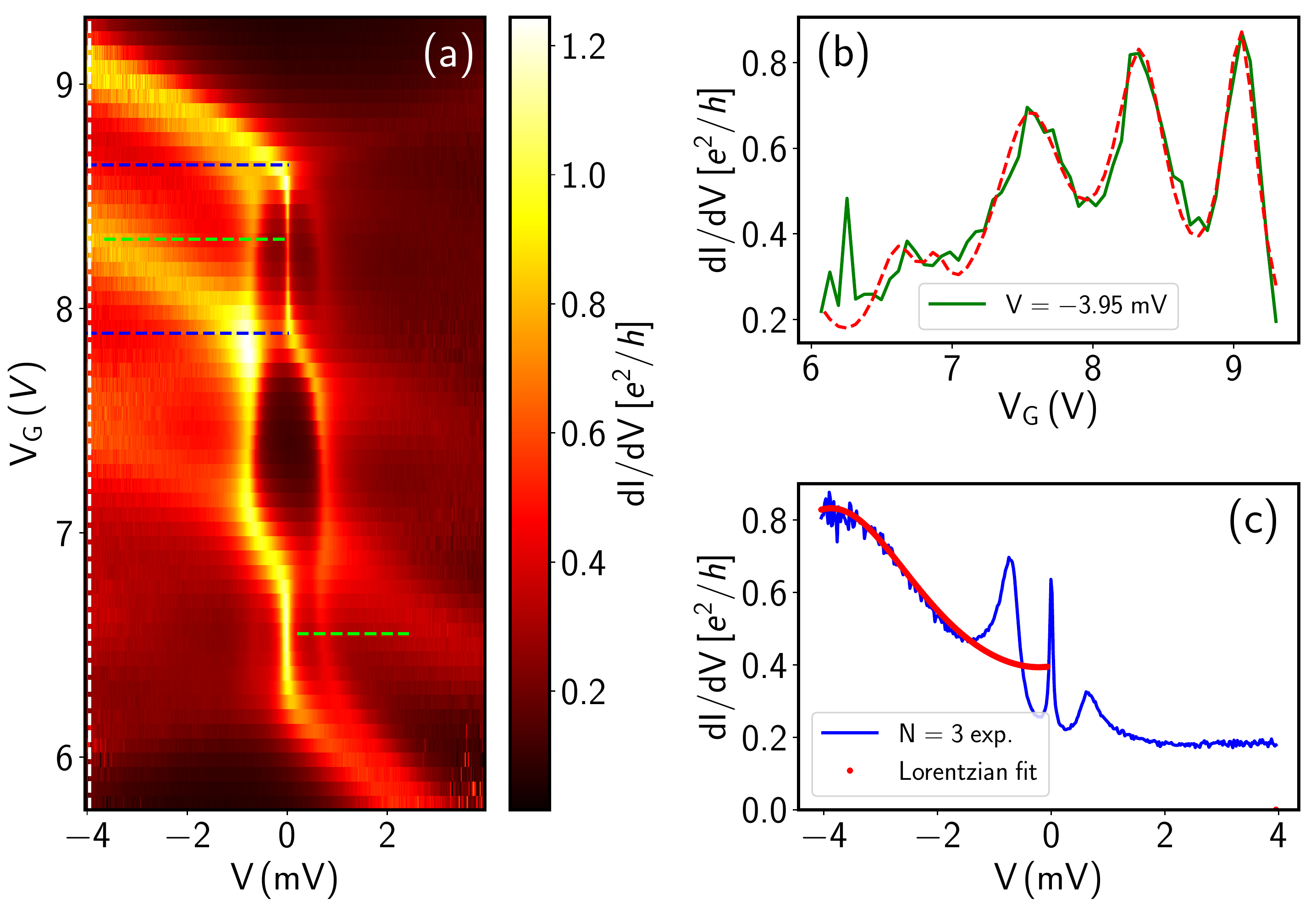}
\caption{ Extracting the contribution from the charge transfer peaks. (a) Experimentally obtained $dI/dV$ data.  The green lines in valley $N=1$, $N=3$  mark the bias required to reach the border of the Coulomb diamond from the middle of the valley; the distance between the blue lines yields the addition energy corrected by the level arm $\alpha_{\rm g}$. (b) Lorentzian fit (red) of the experimental gate trace (green) at $V=-3.95$~mV. 
(c) Contribution to the differential conductance according to the Lorentzian fit (red dashed line) and experimental data (blue) as a function of the bias voltage.  }
\label{keacd}
\end{figure}

We focus here on the resonant lines forming the borders of the Coulomb diamonds. 
Each of the lines corresponds to the electrochemical potential of the dot $\mu(N,V_{\rm g})$  being in resonance with either the source or the drain chemical potentials $\mu_L$, $\mu_R$.
Thus, bias asymmetries impact the slopes of the lines. The strength of the resonance is instead governed by the coupling of the dot levels to the leads.  
From Fig.~\ref{keacd}(a), it is clear that either the bottom-right ($N=1$ valley), or the top-left  ($N=3$ valley) edges harbor the strongest resonant lines.
These lines correspond to the condition $\mu(1,V_{\rm g})=\mu_R$ and $\mu(3,V_{\rm g})=\mu_R$ and suggest a stronger coupling to the right lead.  
%

To extract the tunneling couplings we observe that 
 we can fit gate traces near the resonance using 
a Breit-Wigner conductance formula which is applicable to Coulomb oscillations in the strong-coupling regime~\cite{PhysRevB.44.1646}. We hence approximate the differential conductance with respect to the gate voltage $V_{\rm g}$ by a sum of weighted Lorentzians, each corresponding to a Coulomb oscillation peak:
	\begin{align}
\frac{dI}{dV}\bigg\rvert_{\text{fit}}(V,V_g)&=\frac{e^2}{h}\sum_{l} m_l \Gamma_l\frac{\Gamma_l}{\alpha^2_{\rm g}e^2(V_{\rm g}-c_l)^2+\left(\frac{\Gamma_l}{2}\right)^2}.\label{lorfiteq}
\end{align}
Here $m_l$, $c_l$, and $\Gamma_l$ are the weight, center, and FWHM of the Lorentzian fit of the $l^{th}$ peak. Finally, $\alpha_{\rm g}$ is the level arm accounting for the capacitive coupling to the gate voltage.  The latter can be extracted by inspecting the $N=3$ diamond of Fig.~\ref{keacd}(a). We start by estimating the charging energy $U$, which is given by the length of the green line in the $N=3$ diamond. We obtain $\Delta V=U/e=3.74$~mV.
Likewise the separation $\Delta V_{\rm g}$ of the blue lines bordering the $N=3$ diamond  gives the charging energy scaled by the gate capacitance factor $\alpha_{\rm g}$. Hence we obtain $\alpha_ {\rm g}=\frac{3.74~{\rm mV}}{0.75~{\rm V} }=0.005$. 
Note that  for a good fit of the  gate trace to the data the  low-bias regime must be avoided, to minimize contributions from the inelastic Kondo peaks.  \\
\subsubsection{N = 3}
\label{CNTdotparamN3}
For the top-left border of Fig.~\ref{keacd}(a) ($N=3$ valley) the analytical fitting of the experimental differential conductance is dominated by three Lorentzians originating from the   resonances at the right lead, and is shown in Fig.~\ref{keacd}(b). The fitting parameters of the three dominant peaks are  tabulated in Tab.~\ref{lorfitn3}. In particular, the second lorentzian has $\Gamma_2=2.444$~meV. Since it is this charge transfer peak which most impacts the Kondo resonance, we approximate $\Gamma \approx   \Gamma_2$. Together with $T_{K3}=0.37$~K we find  $\frac{k_{\rm B}T_{K3}}{\Gamma}=0.0131$. Finally, fixing $\frac{W}{\Gamma}=100$, we get $\mu_0-\varepsilon_d=5.3643 ~\Gamma$ using the procedure described in Sec.~\ref{dotlevelcalc}.
 As shown in Fig.~\ref{keacd}(c), the parameters extracted from the fitting can be used to evaluate the contribution to the  differential conductance from the Coulomb peaks also as a function of the bias voltage.  We notice that such contribution is quite remarkable and partly accounts for the different height of the inelastic Kondo peaks. 
%
%
%
%
\begin{table}
\begin{center}
 \begin{tabular}{||c c c c||} 
 \hline
 Peak $(l)$ & $\Gamma_l$~(meV) & {$m_l$} \qquad& $c_l { (V)}$ \\ [0.5ex] 
 \hline\hline
 1 & 1.54 & 0.059 \qquad & 9.06 \\ 
 \hline
 2 & 2.44 & 0.085 \qquad & 8.35 \\
 \hline
 3 & 3.5 & 0.11 \qquad & 7.56 \\
 \hline
\end{tabular}
\end{center}
\caption{ Parameter set used for the fit of the three dominant peaks of the differential conductance for gate voltage ranges relevant for the $N=3$ valley as shown in Fig.~\ref{keacd}(b).  }
\label{lorfitn3}
\end{table}

\subsubsection{N = 1}
\label{CNTdotparamN1}
%
\begin{figure}[h!]
  \includegraphics[width=1\linewidth]{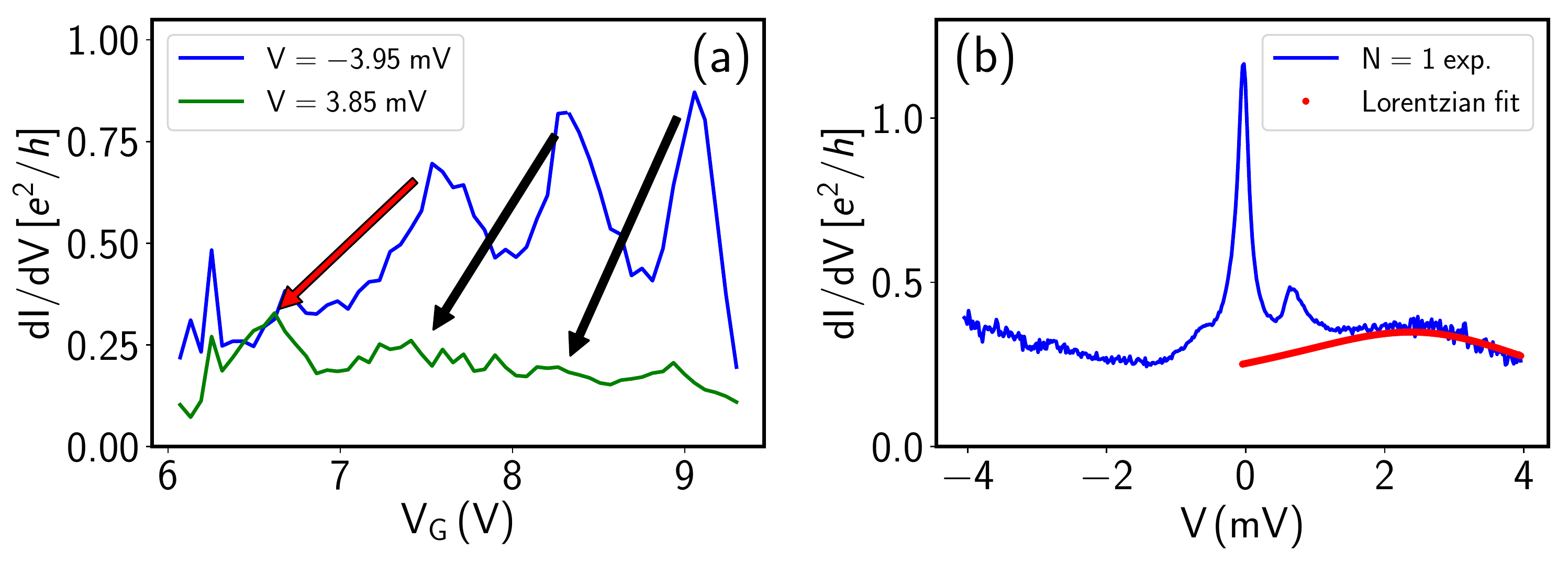}
\caption{Lorentzian fit to the data for the $N=1$ gate range. (a) Experimentally measured  $dI/dV$    at $V_{}=3.85$~mV (green) and $V=-3.95$~mV (blue). The Lorentzian fits to the experimental gate trace at positive bias are shown in red. The arrow marks the  evolution of the  charge-transfer peaks as the gate and bias voltages are varied. Notice that the peak belonging to the lower-left edge of the $N=2$ Coulomb diamond  evolves in the one  belonging the upper-right edge of the $N=1$ Coulomb diamond (as marked by the red arrow). (b) The fitting parameters found in (a) are used in Eq.~\eqref{lorfiteq} and extrapolated to find the differential conductance in the middle of the $N=1$ Coulomb diamond over a range of bias voltages. }
\label{ctn1}
\end{figure}

The lorentzian peak fitting is described and performed for the $N=1$ case in Fig.~\ref{ctn1}. 
The resonance line yielding  the upper right edge of the $N=1$ diamond in Fig.~\ref{keacd}(a), corresponding to $\mu(2)=\mu_R$, is the same line which yields the peak indexed $l=3$ in Tab.~\ref{lorfitn3} at positive bias voltages. The line has the same linewidth $\Gamma_3=3.491$~meV but  a diminished magnitude $m_3=0.06$. Hence, for the $N=1$ case approximating $\Gamma\approx\Gamma_3$ and with $T_{K1}=1$~K we find $\frac{k_{\rm B}T_{K1}}{\Gamma}=0.02487$. We choose $\frac{W}{\Gamma}=100$, from which we get $\mu_0-\varepsilon_d=4.8389 \;\Gamma$ and $f(\Delta)=0.4$ following the procedure described in Sec.~\ref{dotlevelcalc}.
 Further,  the Lorentzian fit can be used to estimate the impact of the Coulomb peaks on the differential conductance  vs. bias, as seen in Fig.~\ref{ctn1}(c). Notice that the fit cannot be extended all the way down to zero bias and beyond to the negative bias region as its validity is restricted to the edges of the Coulomb diamonds.


\subsection{Comparison: KEA theory and experiment}
\subsubsection{N=1}
\begin{figure}[h!]
 \includegraphics[width=\linewidth]{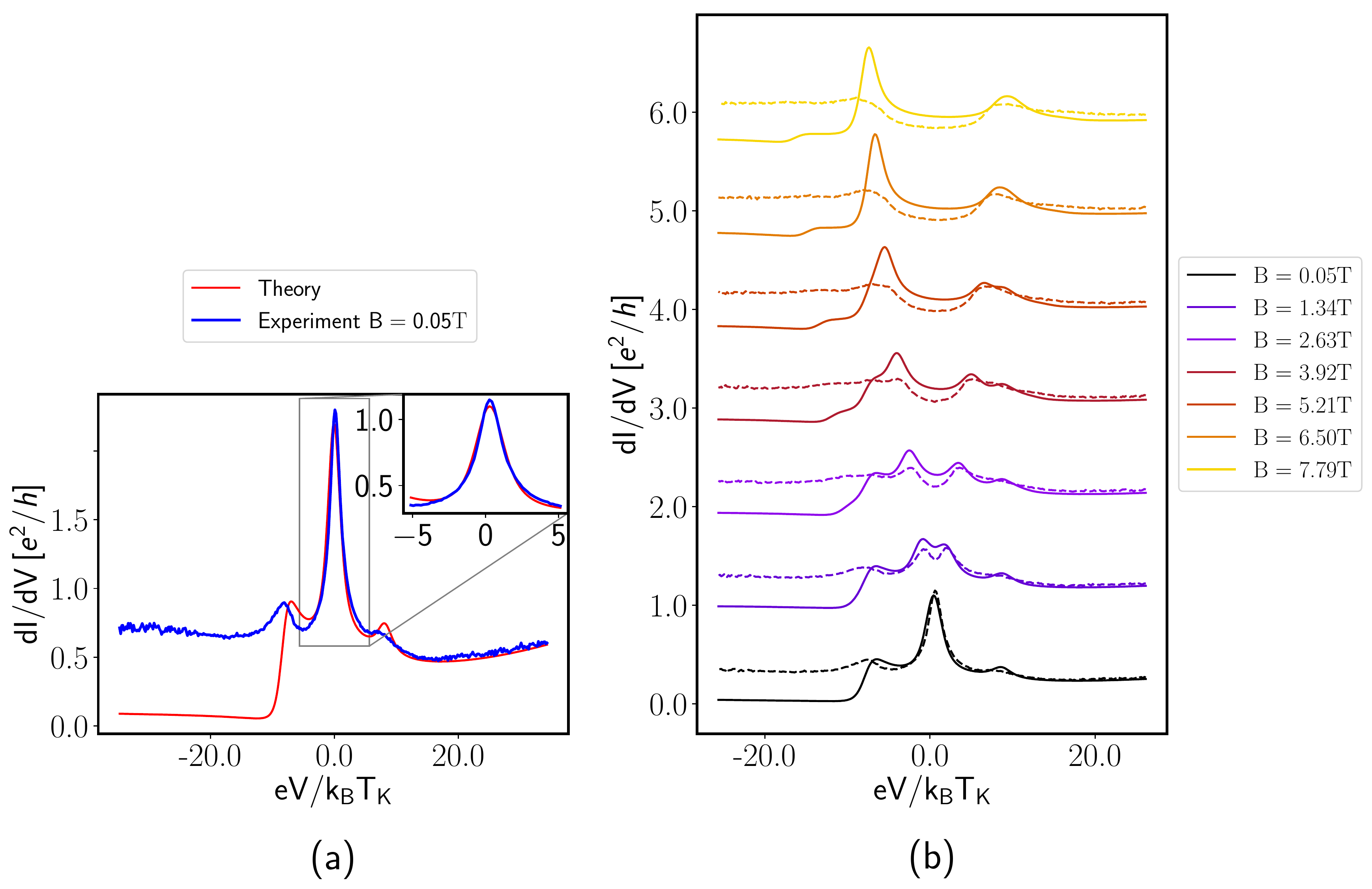}
\caption{Differential conductance for the $N=1$ case, at (a) $B=0.05$~T and (b) for various values of the magnetic field. The blue curve in (a) shows the experimental trace while the red one shows the results of the KEA theory; the inset zooms into the low-bias section. The mismatch between theory and experiment at negative potential drop $eV$ is partly due to the absence of the contribution from the $\mu_R=\mu(2)$ Coulomb peak, which is not included in the  KEA theory  due to the assumption of an infinite charging energy. }
\label{n1thexp}
\end{figure}
The KEA results are compared with the experiment for the $N=1$ case in Fig.~\ref{n1thexp}. 
Since the KEA assumes an infinite charging energy $U$, it cannot account for the contribution from the Coulomb line corresponding to the $\mu(2)=\mu_R$ resonance. Thus, the KEA calculation should be complemented with the lorentzian fit discussed in the previous subsection, being of relevance at positive bias (or negative potential drop). On the other hand the KEA theory fully accounts for the Coulomb lines $\mu_\alpha=\mu(1)$ and hence 
properly describes the negative bias (positive potential drop) region. We notice that while the agreement between theory and experiment is reasonably good at small magnetic fields, it deteriorates as the field is increased.  
 Also, the experimental data on that side is remarkably clear compared to the $N=3$ case, where it is plagued by strong charge-transfer peak tails from the upper edges of the Coulomb diamond and other background conductance which is unaccounted for. In the positive bias region, the charge-transfer peak from the upper right edge of the $N=1$ diamond leads to significant deviations between the experiment and our theory. 

\subsubsection{N=3}

\begin{figure}[h!]
   \includegraphics[width=\linewidth]{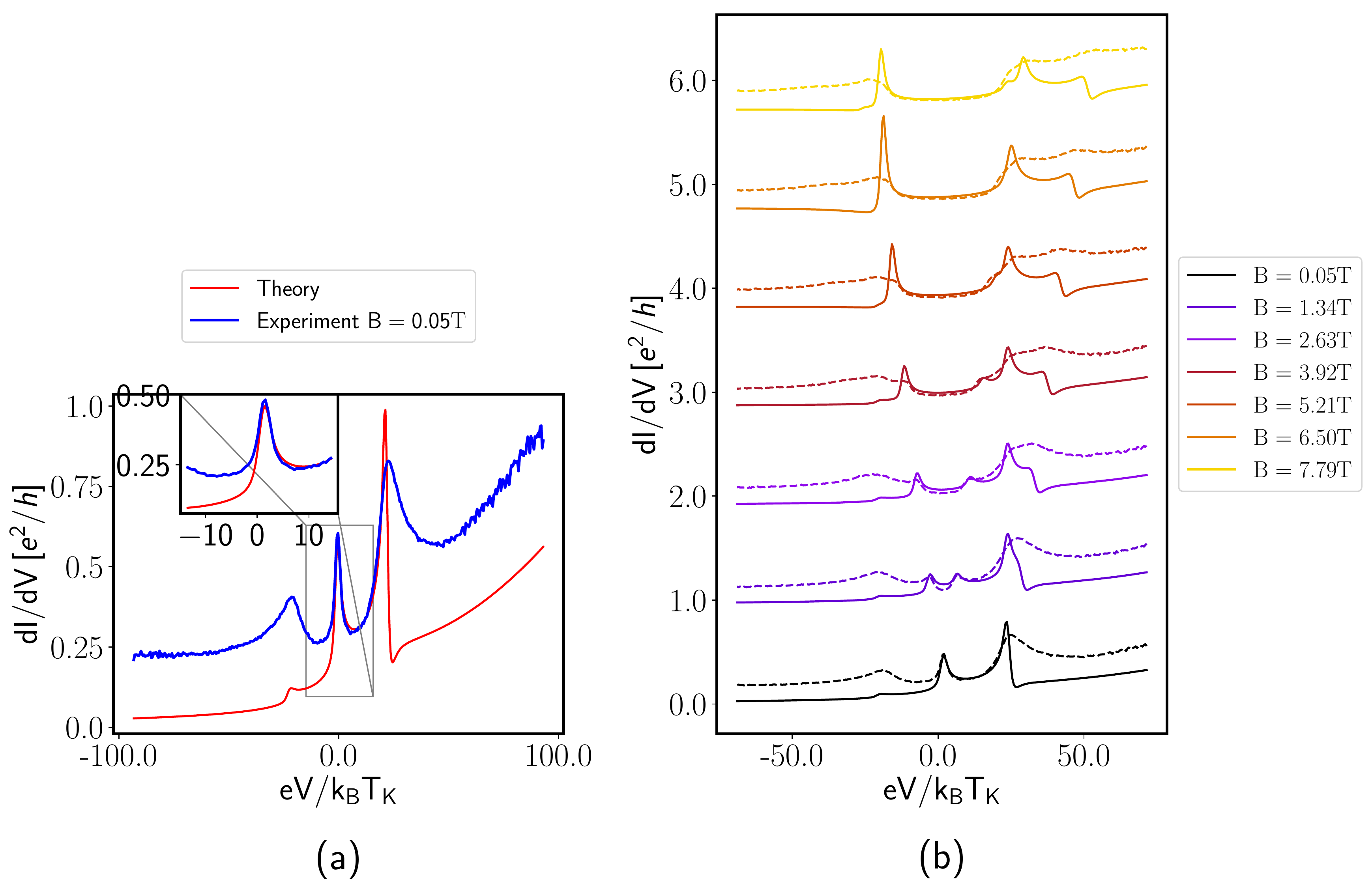}
\caption{Differential conductance for the $N=3$ case, at (a) $B=0.05$~T and (b) for various values of the magnetic field. The blue curve in (a) shows the experimental data while the red curve is the result of the KEA theory; the inset zooms into the low-bias section. The discrepancy at positive bias drop is partly due to the absence of the $\mu(3)=\mu_R$ Coulomb peak in the KEA theory due to the assumption of infinite charging energy.}
\label{n3thexp}
\end{figure}

For the $N=3$ case, the KEA results are matched with the experiments in Fig.~\ref{n3thexp}. The low-bias match is reasonably good, as seen in the inset in Fig.~\ref{n3thexp}(a).
However, the high bias behavior shows a significant deviation between the experiment and the KEA theory. This may partly be attributed to the strong charge-transfer peak on the lower left edge of the $N=3$ Coulomb diamond and a background conductance as seen in Fig.~\ref{keacd}(a).
However, the location of the peaks of the differential conductance are well described by the KEA also at large magnetic fields.  

\bibliographystyle{apsrev}
\bibliography{bibl1}